\documentclass[12pt]{article}
\usepackage{amsfonts,graphicx,amssymb,amsmath,amsthm,epsfig}

\allowdisplaybreaks

\begin{document}
\title{\bf Cosmological Solutions through Gravitational Decoupling in
$f(\mathcal{R},\mathcal{T},\mathcal{R}_{\mathrm{a}\mathrm{b}}\mathcal{T}^{\mathrm{a}\mathrm{b}})$
Gravity}
\author{M. Sharif$^1$ \thanks{msharif.math@pu.edu.pk} and Tayyab Naseer$^{1,2}$ \thanks{tayyabnaseer48@yahoo.com}\\
$^1$ Department of Mathematics and Statistics, The University of Lahore,\\
1-KM Defence Road Lahore, Pakistan.\\
$^2$ Department of Mathematics, University of the Punjab,\\
Quaid-e-Azam Campus, Lahore-54590, Pakistan.}

\date{}
\maketitle
\begin{abstract}
In this paper, we adopt minimal gravitational decoupling scheme to
extend a non-static spherically symmetric isotropic composition to
anisotropic interior in
$f(\mathcal{R},\mathcal{T},\mathcal{R}_{\mathrm{a}\mathrm{b}}\mathcal{T}^{\mathrm{a}\mathrm{b}})$
theory. A geometric deformation is applied only on $g_{rr}$ metric
component through which the modified field equations are separated
into two sets, each of them correspond to their parent (seed and
newly added) source. An isotropic model suggested by the
Friedmann-Lemaitre-Robertson-Walker metric is adopted to reduce the
unknowns in the first set. We then obtain an isotropic solution by
making use of a linear equation of state and a particular form of
the scale factor. A density-like constraint is chosen to solve the
other sector containing the deformation function and multiple
components of an additional matter source. Further, the graphical
interpretation of the developed model is carried out to analyze how
a decoupling parameter and modified gravity influence the
evolutionary phases of the universe. It is concluded that only the
radiation-dominated era meets stability criteria everywhere in this
matter-geometry coupled theory.
\end{abstract}
{\bf Keywords:}
$f(\mathcal{R},\mathcal{T},\mathcal{R}_{\mathrm{a}\mathrm{b}}\mathcal{T}^{\mathrm{a}\mathrm{b}})$
theory;
Cosmology; Gravitational decoupling. \\
{\bf PACS:} 04.40.Dg; 04.50.Kd; 98.80.Jk.

\section{Introduction}

Supernova type Ia \cite{1}, Wilkinson Microwave Anisotropy Probe
\cite{1a} and ejection of x-rays \cite{1b} from galactic clusters
are the cosmological observations that support accelerating
expansion of the universe. Such expansion is caused by the abundance
of a mysterious component in the universe (i.e., dark energy) that
exerts a tremendous repulsive force on self-gravitating bodies. This
unknown force motivated astronomers to unveil its hidden properties.
In this perspective, researchers made certain geometric corrections
to the Einstein-Hilbert action, leading to a variety of modified
gravitational theories of general relativity ($\mathbb{GR}$). The
$f(\mathcal{R})$ theory is the immediate modification obtained by
substituting the generic function of the Ricci scalar in place of
$\mathcal{R}$ in the Einstein-Hilbert action. This theory has been
studied as the first attempt to explore different evolutionary
epochs, including inflationary and current rapid expansion eras
\cite{2,2a}. Several approaches have been utilized to get some
feasible results corresponding to different $f(\mathcal{R})$ models
\cite{9}-\cite{9h}.

Bertolami et al. \cite{10} proposed an initial idea of the
interaction between matter and geometry in $f(\mathcal{R})$
scenario. They were the first to use the Lagrangian in terms of
$\mathcal{R}$ and $\pounds_{\mathcal{M}}$ to investigate the effects
of such coupling on massive objects. Harko et al. \cite{20} recently
proposed $f(\mathcal{R},\mathcal{T})$ theory by generalizing this
coupling at action level, where $\mathcal{T}$ represents trace of
the energy-momentum tensor ($\mathbb{EMT}$). Baffou et al.
\cite{22a} employed two distinct solutions and confirmed the
stability of this theory in both cases. As far as the cosmological
dynamics is concerned, the corresponding equations have been solved
for high as well as low-redshift regimes and found those solutions
in agreement with the observational results. We have merged the idea
of gravitational decoupling and the complexity factor, and obtained
feasible results in this framework \cite{22ab}. However, such type
of theories do not possess the most general Lagrangian to describe
the non-minimal matter-geometry coupling. Haghani \cite{22}
generalized this theory of gravity by introducing a term
$\mathcal{Q}\equiv\mathcal{R}_{\mathrm{a}\mathrm{b}}\mathcal{T}^{\mathrm{a}\mathrm{b}}$
in the functional, termed as the
$f(\mathcal{R},\mathcal{T},\mathcal{Q})$ theory. A fascinating
difference in $f(\mathcal{R},\mathcal{T})$ theory and in an
insertion of the above term is that the field equations of the
former theory reduce to those of $f(\mathcal{R})$ gravity when one
considers a traceless $\mathbb{EMT}$, i.e., $\mathcal{T}=0$.
However, the appearance of the term
$\mathcal{R}_{\mathrm{a}\mathrm{b}}\mathcal{T}^{\mathrm{a}\mathrm{b}}$
still entails a strong non-minimal coupling to the gravitational
field \cite{22,23}. The conservation of these theories is usually
disappeared that gives rise to the existence of an extra force in
the gravitating field. Consequently, the geodesic motion of test
particles is no more preserved.

Odintsov and S\'{a}ez-G\'{o}mez \cite{23} discussed the occurrence
of instability of the fluid in Friedmann-Lemaitre-Robertson-Walker
(FLRW) solution for different modified models. They also formulated
de Sitter solutions and re-established this theory's action. Sharif
and Zubair \cite{22b} studied thermodynamical laws of the black hole
corresponding to $\mathcal{R}+\gamma\mathcal{Q}$ and
$\mathcal{R}(1+\gamma\mathcal{Q})$ for different choices of
$\pounds_{\mathcal{M}}$. They also dealt with the energy bounds for
these models, deriving multiple possible values of the coupling
constant $\gamma$ for which those constraints are satisfied
\cite{22c}. In the context of $\mathcal{R}+\gamma\mathcal{Q}$
gravity, we have developed solutions to the field equations
representing quark stars and analyzed their acceptability by
graphically interpreting the state determinants and stability
criteria \cite{21}. A non-static self-gravitating system has also
been studied through structure scalars and evolutionary patterns in
the absence and presence of electric charge \cite{21b}.

Cosmological principle declares that the universe is homogeneous and
isotropic on scales greater than $300h^{-1}$ Mpc \cite{9a}. To
describe the expanding behavior of the universe, people commonly use
the FLRW metric, which is consistent with this principle. Numerous
cosmological studies over the last two decades led to the conclusion
that our universe is not isotropic \cite{10aaaa}. For instance, a
small deviation from being isotropic has been observed during the
investigation of inhomogeneous Supernova Ia \cite{10aaa,10aaaaa}.
There are several other inconsistencies (i.e., radio sources
\cite{10aa}, infrared galaxies \cite{10ab} and gamma-ray bursts
\cite{10ac}, etc.) that have been found through various
observations. The stellar structures emit x-rays whose
direction-dependent behavior helped Migkas and Reiprich \cite{10a}
to examine the universe's isotropy. A similar strategy was then
implemented on some other galactic clusters, leading to the
discovery that the universe possesses anisotropy \cite{10b}. This
emphasizes the significance of the development of anisotropic
solutions that would ultimately help astronomers to proper
understand the universe's evolutionary stages.

A self-gravitating system can fully be characterized by the set of
corresponding field equations, and finding their solution is a
challenging task due to the engagement of higher order terms
\cite{10bc}. Gravitational decoupling is the most promising
technique that transforms isotropic configurations into anisotropic
setup through the insertion of an extra gravitating source. This
scheme deforms one or both metric potentials, resulting in two
distinct systems of field equations that represent their parent
sources. Solving both sets independently and then adding them
through a particular relation leads to the solution corresponding to
the total matter configuration. This strategy was recently pioneered
by Ovalle \cite{11} in the braneworld, where he used minimal
geometric deformation (MGD) which permits only $g_{rr}$ component to
deform and formulated exact anisotropic solutions describing
spherical spacetime. Following this, Ovalle et al. \cite{13}
developed two anisotropic extensions to an isotropic geometry in
$\mathbb{GR}$ and examined how the decoupling parameter influences
anisotropy. Recently, Gabbanelli et al. \cite{14} determined
anisotropic Durgapal-Fuloria solution with the help of the same
procedure.

The isotropic configurations characterized by different isotropic
models (such as Heintzmann \cite{17}, Krori-Barua \cite{15} and
Tolman VII \cite{19}) have been extended to their respective
anisotropic solutions. This concept has been extended to different
modified theories by Sharif and his collaborators, where they have
developed multiple anisotropic models from isotropic solutions and
found them physically acceptable for certain parametric values
\cite{16}. This scheme was also followed by Cede{\~n}o and Contreas
\cite{19a} to find anisotropic cosmological Kantowski-Sachs and FLRW
solutions. Various authors have formulated anisotropic
Korkina-Orlyanskii \cite{16b} as well as Tolman VII solutions
\cite{16c} in $f(\mathcal{R},\mathcal{T})$ theory and found viable
and stable results. We have used minimal and extended
transformations on the metric potentials and developed multiple
charged/uncharged acceptable solutions in
$\mathcal{R}+\gamma\mathcal{Q}$ theory \cite{21a}. The isotropic
FLRW model has been extended to the anisotropic universe by Sharif
and Majid \cite{16a} in Brans-Dicke scenario. They also discussed
different evolutionary eras through the developed solution by
choosing a linear equation of state ($\mathbb{E}o\mathbb{S}$).

This paper intends to use MGD strategy to generalize isotropic FLRW
solution to anisotropic domain in the framework of
$f(\mathcal{R},\mathcal{T},\mathcal{R}_{\mathrm{a}\mathrm{b}}\mathcal{T}^{\mathrm{a}\mathrm{b}})$
theory. The paper follows the pattern described below. Section
\textbf{2} is concerned with some fundamental concepts of this
modified theory and the incorporation of an additional gravitating
source in the seed isotropic matter distribution. In section
\textbf{3}, we use a peculiar transformation to decouple the field
equations in two standard sets that are associated with their
originating configurations. The formulated solution is then
interpreted graphically to explore the impact of modified gravity in
section \textbf{4}. All of our findings are summarized in section
\textbf{5}.

\section{$f(\mathcal{R},\mathcal{T},\mathcal{R}_{\mathrm{a}\mathrm{b}}\mathcal{T}^{\mathrm{a}\mathrm{b}})$ Formalism}

The inclusion of generic functional of $\mathcal{R},~\mathcal{T}$
and
$\mathcal{R}_{\mathrm{a}\mathrm{b}}\mathcal{T}^{\mathrm{a}\mathrm{b}}$
in place of $\mathcal{R}$ in the Einstein-Hilbert takes the form
\cite{22}
\begin{align}\label{1}
\mathcal{A}=\int\sqrt{-g}\bigg\{\frac{f(\mathcal{R},\mathcal{T},
\mathcal{R}_{\mathrm{a}\mathrm{b}}\mathcal{T}^{\mathrm{a}\mathrm{b}})}
{16\pi}\bigg\}d^{4}x+\int\sqrt{-g}\pounds_{\mathcal{M}}d^{4}x+\varpi\int\sqrt{-g}\pounds_{\mathcal{Y}}d^{4}x,
\end{align}
where the Lagrangian densities of the seed and additional sources
(gravitationally coupled with the original matter configuration
through the decoupling parameter $\varpi$) are indicated by
$\pounds_{\mathcal{M}}$ and $\pounds_{\mathcal{Y}}$, respectively.
Also, $g$ symbolizes determinant of the metric tensor
($g_{\mathrm{a}\mathrm{b}}$). The field equations in modified
scenario can be obtained by applying variational principle on the
action \eqref{1} as follows
\begin{equation}\label{2}
\mathcal{G}_{\mathrm{a}\mathrm{b}}=8\pi
\mathcal{T}_{\mathrm{a}\mathrm{b}}^{(tot)},
\end{equation}
where
$\mathcal{G}_{\mathrm{a}\mathrm{b}}=\mathcal{R}_{\mathrm{a}\mathrm{b}}-\frac{1}{2}\mathcal{R}g_{\mathrm{a}\mathrm{b}}$
is known as the Einstein tensor that expresses the geometrical
structure. On the other hand, the right side reveals the nature of
matter content which is further classified as
\begin{equation}\label{3}
\mathcal{T}_{\mathrm{a}\mathrm{b}}^{(tot)}=\mathcal{T}_{\mathrm{a}\mathrm{b}}^{(eff)}+\varpi
\mathcal{Y}_{\mathrm{a}\mathrm{b}}=\frac{\mathcal{T}_{\mathrm{a}\mathrm{b}}}{f_{\mathcal{R}}-\pounds_{\mathcal{M}}f_{\mathcal{Q}}}
+\mathcal{T}_{\mathrm{a}\mathrm{b}}^{(D)}+\varpi\mathcal{Y}_{\mathrm{a}\mathrm{b}}.
\end{equation}
It will be discussed later that the newly added source
($\mathcal{Y}_{\mathrm{a}\mathrm{b}}$) makes the considered setup
anisotropic whose impact on the self-gravitating body is studied
with the help of a parameter $\varpi$. Here,
$\mathcal{T}_{\mathrm{a}\mathrm{b}}^{(D)}$ appears due to the
modification in the Einstein-Hilbert action whose value is given as
\begin{align}\nonumber
\mathcal{T}_{\mathrm{a}\mathrm{b}}^{(D)}&=\frac{1}{8\pi\big(f_{\mathcal{R}}-\pounds_{\mathcal{M}}f_{\mathcal{Q}}\big)}
\bigg[\bigg\{f_{\mathcal{T}}+\frac{1}{2}\mathcal{R}f_{\mathcal{Q}}\bigg\}\mathcal{T}_{\mathrm{a}\mathrm{b}}
+\bigg\{\frac{\mathcal{R}}{2}\bigg(\frac{f}{\mathcal{R}}-f_{\mathcal{R}}\bigg)
-\pounds_{\mathcal{M}}f_{\mathcal{T}}\\\nonumber
&-\frac{1}{2}\nabla_{\mathrm{m}}\nabla_{\mathrm{n}}(f_{\mathcal{Q}}\mathcal{T}^{\mathrm{m}\mathrm{n}})\bigg\}
g_{\mathrm{a}\mathrm{b}}-\frac{1}{2}\Box\big\{f_{\mathcal{Q}}\mathcal{T}_{\mathrm{a}\mathrm{b}}\big\}
-\big\{g_{\mathrm{a}\mathrm{b}}\Box-\nabla_{\mathrm{a}}\nabla_{\mathrm{b}}\big\}f_{\mathcal{R}}\\\label{4}
&-2f_{\mathcal{Q}}\mathcal{R}_{\mathrm{m}(\mathrm{a}}\mathcal{T}_{\mathrm{b})}^{\mathrm{m}}
+\nabla_{\mathrm{m}}\nabla_{(\mathrm{a}}\big\{\mathcal{T}_{\mathrm{b})}^{\mathrm{m}}f_{\mathcal{Q}}\big\}
+2\big\{f_{\mathcal{Q}}\mathcal{R}^{\mathrm{m}\mathrm{n}}+f_{\mathcal{T}}g^{\mathrm{m}\mathrm{n}}\big\}\frac{\partial^2
\pounds_{\mathcal{M}}}{\partial g^{\mathrm{a}\mathrm{b}}\partial
g^{\mathrm{m}\mathrm{n}}}\bigg],
\end{align}
and take $\pounds_\mathcal{M}=-\mu$ leading to $\frac{\partial^2
\pounds_{\mathcal{M}}}{\partial g^{\mathrm{a}\mathrm{b}}\partial
g^{\mathrm{m}\mathrm{n}}}=0$. Here, $f_{\mathcal{R}}=\frac{\partial
f(\mathcal{R},\mathcal{T},\mathcal{Q})}{\partial
\mathcal{R}},~f_{\mathcal{T}}=\frac{\partial
f(\mathcal{R},\mathcal{T},\mathcal{Q})}{\partial
\mathcal{T}},~f_{\mathcal{Q}}=\frac{\partial
f(\mathcal{R},\mathcal{T},\mathcal{Q})}{\partial \mathcal{Q}}$ and
$\nabla_\mathrm{b}$ is the covariant derivative. Moreover,
$\Box\equiv\frac{1}{\sqrt{-g}}\partial_\mathrm{a}\big(\sqrt{-g}g^{\mathrm{a}\mathrm{b}}\partial_{\mathrm{b}}\big)$
denotes the D'Alembert operator. The following equation provides the
trace of modified field equations as
\begin{align}\nonumber
&3\nabla^{\mathrm{n}}\nabla_{\mathrm{n}}f_\mathcal{R}-\mathcal{R}\left(\frac{\mathcal{T}}{2}f_\mathcal{Q}-f_\mathcal{R}\right)
-\mathcal{T}(8\pi+f_\mathcal{T})+\frac{1}{2}\nabla^{\mathrm{n}}\nabla_{\mathrm{n}}(f_\mathcal{Q}\mathcal{T})\\\nonumber
&+\nabla_\mathrm{a}\nabla_\mathrm{b}(f_\mathcal{Q}\mathcal{T}^{\mathrm{a}\mathrm{b}})-2f+(\mathcal{R}f_\mathcal{Q}
+4f_\mathcal{T})\pounds_\mathcal{M}+2\mathcal{R}_{\mathrm{a}\mathrm{b}}\mathcal{T}^{\mathrm{a}\mathrm{b}}f_\mathcal{Q}\\\nonumber
&-2g^{\mathrm{b}\mathrm{n}}
\frac{\partial^2\pounds_\mathcal{M}}{\partial
g^{\mathrm{b}\mathrm{n}}\partial
g^{\mathrm{a}\mathrm{m}}}\left(f_\mathcal{T}g^{\mathrm{a}\mathrm{m}}+f_\mathcal{Q}\mathcal{R}^{\mathrm{a}\mathrm{m}}\right)=0.
\end{align}
We consider isotropic fluid as a seed source whose $\mathbb{EMT}$
can be expressed as
\begin{equation}\label{5}
\mathcal{T}_{\mathrm{a}\mathrm{b}}=(\mu+P)\mathrm{K}_{\mathrm{a}}\mathrm{K}_{\mathrm{b}}+Pg_{\mathrm{a}\mathrm{b}},
\end{equation}
where $P,~\mu$ and $\mathrm{K}_{\mathrm{b}}$ indicate the pressure,
energy density and the four-velocity, respectively.

The inner region of a non-static spherically symmetric
self-gravitating spacetime is defined by the following line element
\begin{equation}\label{6}
ds^{2}=-e^{\alpha_{1}}dt^{2}+e^{\alpha_{2}}dr^{2}+\mathcal{C}^{2}(d\theta^{2}+{\sin^{2}\theta}{d\phi^2}),
\end{equation}
where $\alpha_{1}=\alpha_{1}(t,r),~\alpha_{2}=\alpha_{2}(t,r)$ and
$\mathcal{C}=\mathcal{C}(t,r)$. The corresponding four-velocity is
given as
\begin{equation}\label{7}
\mathrm{K}_\mathrm{b}=(-e^{\frac{\alpha_{1}}{2}},0,0,0),
\end{equation}
with $\mathrm{K}_{\mathrm{b}}\mathrm{K}^{\mathrm{b}}=-1$. We shall
obtain extended cosmological solutions representing different eras
of our universe. Further, we shall analyze our results by taking a
particular
$f(\mathcal{R},\mathcal{T},\mathcal{R}_{\mathrm{a}\mathrm{b}}\mathcal{T}^{\mathrm{a}\mathrm{b}})$
model. The linear model in this theory which entails non-minimal
matter-geometry coupling is of the form \cite{22}
\begin{equation}\label{7a}
f(\mathcal{R},\mathcal{T},\mathcal{R}_{\mathrm{a}\mathrm{b}}\mathcal{T}^{\mathrm{a}\mathrm{b}})=f_1(\mathcal{R})+
f_2(\mathcal{T})+f_3(\mathcal{R}_{\mathrm{a}\mathrm{b}}\mathcal{T}^{\mathrm{a}\mathrm{b}})=\mathcal{R}+2\gamma_1\mathcal{T}
+\gamma_2\mathcal{R}_{\mathrm{a}\mathrm{b}}\mathcal{T}^{\mathrm{a}\mathrm{b}},
\end{equation}
where $\gamma_1$ and $\gamma_2$ are real-valued coupling constants
whose different values lead us to analyze the impact of this
modified theory on compact structures. It must be mentioned here
that only the dimensionally correct equations are of researcher's
interest in the scientific community. We observe that the dimension
of $\mathcal{R}$ and $\mathcal{T}$ is $\frac{1}{l^2}$, thus we have
to choose $\gamma_1$ as a dimensionless constant. Furthermore, the
terms $\mathcal{R}_{\mathrm{a}\mathrm{b}}$ and
$\mathcal{T}^{\mathrm{a}\mathrm{b}}$ have dimension $\frac{1}{l^2}$,
thus the right hand side of the above equation would be
dimensionally balanced only if the constant $\gamma_2$ has dimension
$l^2$.

Several acceptable anisotropic solutions have been developed
corresponding to this model along with $\gamma_2=0$ \cite{39}. The
model \eqref{7a} has also been used in studying anisotropic compact
structures for $\gamma_1=0$ \cite{21b,21a}. The solution
corresponding to this model shows an oscillatory behavior for
positive values of $\gamma_2$, while the scale factor has a
hyperbolic (cosine-type) dependence for $\gamma_2<0$ \cite{22}.
Here, the entities $\mathcal{R},~\mathcal{T}$ and $\mathcal{Q}$ are
\begin{align}\nonumber
\mathcal{R}&=\frac{1}{2e^{\alpha_1+\alpha_2}}\bigg\{e^{\alpha_1}\bigg(\alpha'_1\alpha'_2-\alpha_1'^2
-2\alpha''_1+\frac{4\alpha'_2\mathcal{C}'}{\mathcal{C}}-\frac{4\alpha'_1\mathcal{C}'}{\mathcal{C}}
-\frac{8\mathcal{C}''}{\mathcal{C}}-\frac{4\mathcal{C}'^2}{\mathcal{C}^2}\bigg)\\\nonumber
&+e^{\alpha_2}\bigg(\dot{\alpha}_2^2-\dot{\alpha}_1\dot{\alpha}_2+2\ddot{\alpha}_2
+\frac{4\dot{\mathcal{C}}\dot{\alpha}_2}{\mathcal{C}}-\frac{4\dot{\mathcal{C}}\dot{\alpha}_1}{\mathcal{C}}
+\frac{8\ddot{\mathcal{C}}}{\mathcal{C}}+\frac{4\dot{\mathcal{C}}^2}{\mathcal{C}^2}\bigg)\bigg\}
+\frac{2}{\mathcal{C}^2},\\\nonumber
\mathcal{T}&=-\mu+3P,\\\nonumber
\mathcal{Q}&=\frac{1}{4\mathcal{C}\big(e^{\alpha_1+\alpha_2}\big)}\bigg[\mu\big\{e^{\alpha_1}\alpha_1'^2\mathcal{C}
-e^{\alpha_1}\alpha_1'\alpha_2'\mathcal{C}-e^{\alpha_2}\dot{\alpha}_2^2\mathcal{C}
+e^{\alpha_2}\dot{\alpha}_1\dot{\alpha}_2\mathcal{C}+2e^{\alpha_1}\alpha_1''\mathcal{C}\\\nonumber
&+4e^{\alpha_1}\alpha_1'\mathcal{C}'-2e^{\alpha_2}\ddot{\alpha}_2\mathcal{C}+4e^{\alpha_2}\dot{\alpha}_1\dot{\mathcal{C}}
-8e^{\alpha_2}\ddot{\mathcal{C}}\big\}-P\bigg\{e^{\alpha_2}\dot{\alpha}_1\dot{\alpha}_2\mathcal{C}
-e^{\alpha_1}\alpha_1'\alpha_2'\mathcal{C}\\\nonumber
&-8e^{\alpha_2}\dot{\alpha}_2\dot{\mathcal{C}}+4e^{\alpha_1}\alpha_1'\mathcal{C}'+2e^{\alpha_1}\alpha_1''\mathcal{C}
-8e^{\alpha_1}\alpha_2'\mathcal{C}'-2e^{\alpha_2}\ddot{\alpha}_2\mathcal{C}
+8\ddot{\mathcal{C}}\big(e^{\alpha_1}-e^{\alpha_2}\big)\\\nonumber
&-e^{\alpha_2}\dot{\alpha}_2^2\mathcal{C}+e^{\alpha_1}\alpha_1'^2\mathcal{C}+4e^{\alpha_2}\dot{\alpha}_1\dot{\mathcal{C}}
+8e^{\alpha_1}\mathcal{C}''+\frac{8}{\mathcal{C}}\big(e^{\alpha_1}\mathcal{C}'^2-e^{\alpha_2}\dot{\mathcal{C}}^2
-e^{\alpha_1+\alpha_2}\big)\bigg\}\bigg],
\end{align}
where $'=\frac{\partial}{\partial{r}}$ and
$.=\frac{\partial}{\partial{t}}$. The modified field equations for a
sphere \eqref{6} and standard functional \eqref{7a} are of the form
\begin{align}\label{8}
\mathcal{G}_{0}^{0}&=\frac{1}{1+\mu\gamma_2}\big[\big(8\pi+\gamma_1\big)\mu-3P\gamma_1+\gamma_2\mathcal{T}_{0}^{0(D)}\big]
-8\pi\varpi\mathcal{Y}_{0}^{0},\\\label{9}
\mathcal{G}_{1}^{1}&=\frac{1}{1+\mu\gamma_2}\big[\big(8\pi+5\gamma_1\big)P+\mu\gamma_1+\gamma_2\mathcal{T}_{1}^{1(D)}\big]
+8\pi\varpi\mathcal{Y}_{1}^{1},\\\label{9a}
\mathcal{G}_{2}^{2}&=\frac{1}{1+\mu\gamma_2}\big[\big(8\pi+5\gamma_1\big)P+\mu\gamma_1+\gamma_2\mathcal{T}_{2}^{2(D)}\big]
+8\pi\varpi\mathcal{Y}_{2}^{2},\\\label{9b}
\mathcal{G}_{1}^{0}&=\frac{\gamma_2\mathcal{T}_{1}^{0(D)}}{1+\mu\gamma_2}-8\pi\varpi\mathcal{Y}_{1}^{0}.
\end{align}
The factors multiplied with $\gamma_1$ and $\gamma_2$ in the above
equations arise owing to the modified theory. The entities
$\mathcal{T}_{0}^{0(D)},~\mathcal{T}_{1}^{1(D)},~\mathcal{T}_{2}^{2(D)}$
and $\mathcal{T}_{1}^{0(D)}$ are provided in Appendix \textbf{A}.
These equations show anisotropic configured system only if
$\mathcal{Y}_{1}^{1} \neq \mathcal{Y}_{2}^{2}$. Moreover, the
geometric quantities appear in Eqs.\eqref{8}-\eqref{9b} are given as
follows
\begin{align}\nonumber
\mathcal{G}_{0}^{0}&=\frac{1}{\mathcal{C}^2}-\frac{e^{-\alpha_{2}}}{\mathcal{C}}\bigg(\frac{\mathcal{C}^{'2}}{\mathcal{C}}
-\mathcal{C}^{'}\alpha_{2}^{'}+2\mathcal{C}^{''}\bigg)+\frac{e^{-\alpha_{1}}}{\mathcal{C}}\bigg(\frac{\dot{\mathcal{C}}^2}
{\mathcal{C}}+\dot{\mathcal{C}}\dot{\alpha_{2}}\bigg),\\\nonumber
\mathcal{G}_{1}^{1}&=-\frac{1}{\mathcal{C}^2}-\frac{e^{-\alpha_{1}}}{\mathcal{C}}\bigg(\frac{\dot{\mathcal{C}}^2}{\mathcal{C}}
-\dot{\mathcal{C}}\dot{\alpha_{1}}+2\ddot{\mathcal{C}}\bigg)+\frac{e^{-\alpha_{2}}}{\mathcal{C}}\bigg(\frac{\mathcal{C}^{'2}}
{\mathcal{C}}+\mathcal{C}^{'}\alpha_{1}^{'}\bigg),\\\nonumber
\mathcal{G}_{2}^{2}&=e^{-\alpha_{2}}\bigg(\frac{\alpha_{1}^{'2}}{4}-\frac{\alpha_{1}^{'}\alpha_{2}^{'}}{4}
+\frac{\alpha_{1}^{''}}{2}+\frac{\mathcal{C}^{''}}{\mathcal{C}}\bigg)-e^{-\alpha_{1}}\bigg(\frac{\dot{\alpha_{2}}^{2}}{4}
-\frac{\dot{\alpha_{1}}\dot{\alpha_{2}}}{4}+\frac{\ddot{\alpha_{2}}}{2}+\frac{\ddot{\mathcal{C}}}{\mathcal{C}}\bigg)\\\nonumber
&+\frac{1}{2\mathcal{C}}\big\{e^{-\alpha_{2}}\mathcal{C}^{'}\big(\alpha_{1}^{'}-\alpha_{2}^{'}\big)
+e^{-\alpha_{1}}\dot{\mathcal{C}}\big(\dot{\alpha_{1}}-\dot{\alpha_{2}}\big)\big\},\\\nonumber
\mathcal{G}_{1}^{0}&=\frac{e^{-\alpha_{1}}}{\mathcal{C}}\big(2\dot{\mathcal{C}'}-\dot{\mathcal{C}}\alpha_{1}^{'}
-\mathcal{C}^{'}\dot{\alpha_{2}}\big).
\end{align}

The inclusion of the terms $\mathcal{T}$ and $\mathcal{Q}$ in
generic functional results in non-conserved $\mathbb{EMT}$ lead to
\begin{align}\nonumber
\nabla^\mathrm{a}
\big(\mathcal{T}_{\mathrm{a}\mathrm{b}}+\varpi\mathcal{Y}_{\mathrm{a}\mathrm{b}}\big)&=\frac{2}{4\gamma_1+\gamma_2\mathcal{R}
+16\pi}\bigg[\gamma_2\nabla_\mathrm{a}\big(\mathcal{R}^{\mathrm{m}\mathrm{a}}\mathcal{T}_{\mathrm{m}\mathrm{b}}\big)
+\gamma_2\mathcal{G}_{\mathrm{a}\mathrm{b}}\nabla^\mathrm{a}\mu\\\label{g11}
&-2\gamma_1\nabla_\mathrm{b}\mu-\frac{1}{2}\nabla_\mathrm{b}\mathcal{T}^{\mathrm{ma}}\big(2\gamma_1g_{\mathrm{ma}}
+\gamma_2\mathcal{R}_{\mathrm{ma}}\big)
-\frac{\gamma_2}{2}\nabla^{\mathrm{a}}\mathcal{R}\mathcal{T}_{\mathrm{a}\mathrm{b}}\bigg].
\end{align}

\section{Gravitational Decoupling}

This section investigates the implementation of gravitational
decoupling by MGD approach so that we can reduce the degrees of
freedom of highly complicated equations \eqref{8}-\eqref{9b} and
obtain exact solution. The system of field equations is observed to
be under-determined as it encompasses nine unknowns
($\alpha_{1},\alpha_{2},\mathcal{C},\mu,P,\mathcal{Y}^{0}_{0},\mathcal{Y}^{1}_{1},\mathcal{Y}^{2}_{2},\mathcal{Y}^{0}_{1}$),
thus some restrictions are required to impose. Since the seed source
is coupled with additional anisotropic configuration, therefore we
use MGD scheme that separates them into two sets corresponding to
their original sources which will further be solved independently.
In this regard, a particular transformation to decouple such
equations has initially been proposed by Ovalle \cite{11} for the
static geometry. We now adopt the corresponding extension of this
transformation for the non-static scenario as
\begin{equation}\label{16}
e^{-\alpha_{2}(t,r)} \mapsto
e^{-\alpha_{3}(t,r)}\{1+\varpi\bar{s}(t,r)\},
\end{equation}
where $\bar{s}(t,r)$ is treated as the deformation function.

The implementation of the above transformation on
Eqs.\eqref{8}-\eqref{9b} provides the field equations ($\varpi=0$)
representing an isotropic source as
\begin{align}\nonumber
8\pi\mu&=\big(3P-\mu\big)\gamma_1-\gamma_2\mathcal{T}_{0}^{0(D)}+\big(1+\mu\gamma_2\big)\\\label{17}
&\times\bigg[\frac{\dot{\mathcal{C}}e^{-\alpha_{1}}\dot{\alpha_{3}}}{\mathcal{C}}+\frac{\dot{\mathcal{C}}^2
e^{-\alpha_{1}}}{\mathcal{C}^2}+\frac{\mathcal{C}'e^{-\alpha_{3}}\alpha_{3}'}{\mathcal{C}}+\frac{1}{\mathcal{C}^2}
-\frac{\mathcal{C}'^2e^{-\alpha_{3}}}{\mathcal{C}^2}-\frac{2\mathcal{C}''e^{-\alpha_{3}}}{\mathcal{C}}\bigg],\\\nonumber
8\pi{P}&=-\big(5P+\mu\big)\gamma_1-\gamma_2\mathcal{T}_{1}^{1(D)}+\big(1+\mu\gamma_2\big)\\\label{17a}
&\times\bigg[\frac{\mathcal{C}'\alpha_{1}'e^{-\alpha_{3}}}{\mathcal{C}}+\frac{\dot{\mathcal{C}}
e^{-\alpha_{1}}\dot{\alpha_{1}}}{\mathcal{C}}-\frac{\dot{\mathcal{C}}^2e^{-\alpha_{1}}}{\mathcal{C}^2}
-\frac{1}{\mathcal{C}^2}-\frac{2\ddot{\mathcal{C}}e^{-\alpha_{1}}}{\mathcal{C}}
+\frac{\mathcal{C}'^2e^{-\alpha_{3}}}{\mathcal{C}^2}\bigg],\\\nonumber
8\pi{P}&=-\big(5P+\mu\big)\gamma_1-\gamma_2\mathcal{T}_{2}^{2(D)}+\big(1+\mu\gamma_2\big)\\\nonumber
&\times\bigg[e^{-\alpha_{1}}\bigg\{\frac{\dot{\mathcal{C}}\dot{\alpha_{1}}}{2\mathcal{C}}
-\frac{\dot{\mathcal{C}}\dot{\alpha_{3}}}{2\mathcal{C}}-\frac{\ddot{\mathcal{C}}}{\mathcal{C}}
+\frac{1}{4}\dot{\alpha_{1}}\dot{\alpha_{3}}-\frac{1}{4}\dot{\alpha_{3}}^2-\frac{1}{2}\ddot{\alpha_{3}}\bigg\}\\\label{17b}
&+e^{-\alpha_{3}}\bigg\{\frac{\mathcal{C}'\alpha_{1}'}{2\mathcal{C}}-\frac{\mathcal{C}'\alpha_{3}'}{2\mathcal{C}}
+\frac{\mathcal{C}''}{\mathcal{C}}-\frac{1}{4}\alpha_{1}'\alpha_{3}'+\frac{1}{4}\alpha_{1}'^2
+\frac{1}{2}\alpha_{1}''\bigg\}\bigg],\\\label{17c}
0&=\gamma_2\mathcal{T}_{1}^{0(D)}+\big(1+\mu\gamma_2\big)\bigg[\frac{\dot{\mathcal{C}}e^{-\alpha_{1}}\alpha_{1}'}
{\mathcal{C}}+\frac{\mathcal{C}'e^{-\alpha_{1}}\dot{\alpha_{3}}}{\mathcal{C}}
-\frac{2\dot{\mathcal{C}'}e^{-\alpha_{1}}}{\mathcal{C}}\bigg].
\end{align}
The system \eqref{17}-\eqref{17c} still involves one extra degree of
freedom, therefore we need one more constraint to deal with it.
Multiple $\mathbb{E}o\mathbb{S}s$ have been employed in the
literature, one of them is given in the following that linearly
connects the energy density with pressure as
\begin{equation}\label{16a}
P=\gamma_3\mu,
\end{equation}
where certain values of the parameter $\gamma_3$ help to study
different eras. In particular, we have
\begin{itemize}
\item $\gamma_3=\frac{1}{3}$ \quad $\Rightarrow$ \quad radiation-dominated era,
\item $\gamma_3=0$ \quad $\Rightarrow$ \quad matter-dominated era,
\item $\gamma_3=-1$ \quad $\Rightarrow$ \quad vacuum energy dominated era.
\end{itemize}
The explicit expressions for the state determinants such as energy
density and pressure are calculated through Eqs.\eqref{17} and
\eqref{17a} along with $\mathbb{E}o\mathbb{S}$ \eqref{16a} as
\begin{align}\nonumber
\mu&=\big[e^{\alpha_1+\alpha_3}\big\{3 \gamma_1 \gamma_3
\mathcal{C}^2 -\gamma_1 \mathcal{C}^2-8 \pi \mathcal{C}^2+\gamma_2
\big\}+\gamma_2e^{\alpha_3}\big\{ \mathcal{C}\dot{\mathcal{C}}
\dot{\alpha}_3+\dot{\mathcal{C}}^2\big\}\\\nonumber
&+\gamma_2e^{\alpha_1}\big\{ \mathcal{C}' \mathcal{C}
\alpha_3'-2\mathcal{C}\mathcal{C}'' -\mathcal{C}'^2
\big\}\big]^{-1}\big[e^{\alpha_1}\big\{2 \mathcal{C}\mathcal{C}''
-\mathcal{C}\mathcal{C}'\alpha_3'+\mathcal{C}'^2\big\}\\\label{16b}
&-e^{\alpha_3}\big\{\mathcal{C}\dot{\mathcal{C}}
\dot{\alpha}_3+\dot{\mathcal{C}}^2\big\}+e^{\alpha_1+\alpha_3}\big\{\gamma_2\mathcal{T}_{0}^{0(D)}\mathcal{C}^2
-1\big\}\big],\\\nonumber
P&=\gamma_3\big[\gamma_2e^{\alpha_3}\big\{2
\ddot{\mathcal{C}}\mathcal{C} -\mathcal{C}\dot{\mathcal{C}}
\dot{\alpha}_1+ \dot{\mathcal{C}}^2
\big\}-\gamma_2e^{\alpha_1}\big\{ \mathcal{C}\mathcal{C}' \alpha_1'+
\mathcal{C}'^2 \big\}\\\nonumber &+e^{\alpha_1+\alpha_3}\big\{5
\gamma_1\gamma_3\mathcal{C}^2+\gamma_1\mathcal{C}^2+8\pi\gamma_3\mathcal{C}^2+\gamma_2\big\}\big]^{-1}
\big[\mathcal{C}'e^{\alpha_1}\big\{\mathcal{C}\alpha_1'+\mathcal{C}'\big\}\\\label{16c}
&-e^{\alpha_3}\big\{2\mathcal{C}\ddot{\mathcal{C}}+\dot{\mathcal{C}}^2-\mathcal{C}\dot{\mathcal{C}}\dot{\alpha}_1\big\}
-e^{\alpha_1+\alpha_3}\big\{\gamma_2\mathcal{T}_{1}^{1(D)}\mathcal{C}^2+1\big\}\big].
\end{align}
Likewise, another set portraying the anisotropic source is also
obtained from the field equations \eqref{8}-\eqref{9b} for
$\varpi=1$ as
\begin{align}\label{18}
8\pi\mathcal{Y}^{0}_{0}&=\frac{\dot{\mathcal{C}} \dot{\bar{s}}
e^{-\alpha_{1}}}{\mathcal{C} (\varpi\bar{s}+1)}+\frac{\mathcal{C}'
\bar{s}' e^{-\alpha_{3}}}{\mathcal{C}}-\frac{\mathcal{C}' \bar{s}
e^{-\alpha_{3}} \alpha_{3}'}{\mathcal{C}}+\frac{\mathcal{C}'^2
\bar{s} e^{-\alpha_{3}}}{\mathcal{C}^2}+\frac{2 \mathcal{C}''
\bar{s} e^{-\alpha_{3}}}{\mathcal{C}},\\\label{18a}
8\pi\mathcal{Y}^{1}_{1}&=\frac{\mathcal{C}' \bar{s} \alpha_{1}'
e^{-\alpha_{3}}}{\mathcal{C}}+\frac{\mathcal{C}'^2
\bar{s}e^{-\alpha_{3}}}{\mathcal{C}^2},\\\nonumber
8\pi\mathcal{Y}^{2}_{2}&=\frac{\dot{\mathcal{C}} \dot{\bar{s}}
e^{-\alpha_{1}}}{2\mathcal{C} (\varpi\bar{s}+1)}+\frac{\mathcal{C}'
\bar{s}' e^{-\alpha_{3}}}{2 \mathcal{C}}+\frac{\mathcal{C}' \bar{s}
\alpha_{1}' e^{-\alpha_{3}}}{2 \mathcal{C}}-\frac{\mathcal{C}'
\bar{s} e^{-\alpha_{3}} \alpha_{3}'}{2
\mathcal{C}}+\frac{\mathcal{C}'' \bar{s}
e^{-\alpha_{3}}}{\mathcal{C}}\\\nonumber&-\frac{\dot{\bar{s}}
e^{-\alpha_{1}} \dot{\alpha_{1}}}{4
(\varpi\bar{s}+1)}+\frac{\dot{\bar{s}} e^{-\alpha_{1}}
\dot{\alpha_{3}}}{2 (\varpi\bar{s}+1)}-\frac{3 \varpi\dot{\bar{s}}^2
e^{-\alpha_{1}}}{4 (\varpi\bar{s}+1)^2}+\frac{\ddot{\bar{s}}
e^{-\alpha_{1}}}{2 (\varpi\bar{s}+1)}+\frac{1}{4} \bar{s}'
\alpha_{1}' e^{-\alpha_{3}}\\\label{18b}&-\frac{1}{4} \bar{s}
\alpha_{1}' e^{-\alpha_{3}} \alpha_{3}'+\frac{1}{4} \bar{s}
\alpha_{1}'^2 e^{-\alpha_{3}}+\frac{1}{2} \bar{s} \alpha_{1}''
e^{-\alpha_{3}},\\\label{18c}
8\pi\mathcal{Y}^{0}_{1}&=-\frac{\mathcal{C}' \dot{\bar{s}}
e^{-\alpha_{1}}}{\mathcal{C} (\varpi\bar{s}+1)}.
\end{align}

\section{Anisotropic Cosmological Solution}

We know that FLRW metric represents homogeneous universe, thus we
take isotropic matter source that will be extended to the
anisotropic domain. The FLRW metric engaging a scale factor
$\mathrm{a}(t)$ has the form
\begin{equation}\label{32}
ds^{2}=-dt^{2}+\mathrm{a}^2(t)\bigg(\frac{dr^{2}}{1-\mathrm{k}r^2}+r^2d\theta^{2}+r^2{\sin^{2}\theta}{d\phi^2}\bigg),
\end{equation}
where $\mathrm{a}(t)$ quantifies the variation of distance between
different points due to expanding nature of the universe. Also, the
curvature parameter is symbolized by $\mathrm{k}$ whose distinct
values ($0,-1,1$) correspond to flat, open and close models of the
universe, respectively. Equating the metric potentials of the line
elements \eqref{6} and \eqref{32}, we obtain some relations as
follows
\begin{align}\label{33}
e^{\alpha_{1}(t,r)}=1, \quad
e^{\alpha_{3}(t,r)}=\frac{\mathrm{a}^2(t)}{1-\mathrm{k}r^2}, \quad
\mathcal{C}(t,r)=r \mathrm{a}(t).
\end{align}
Equations \eqref{16b} and \eqref{16c} take the form after utilizing
the above relations as
\begin{align}\label{36}
\mu&=\frac{\gamma_2\mathrm{a}^2\mathcal{T}_{0}^{0(D)}-3\left(\dot{\mathrm{a}}^2+\mathrm{k}\right)}
{3\gamma_2(\dot{\mathrm{a}}^2+\mathrm{k})-\mathrm{a}^2\{8\pi+\gamma_1(1-3\gamma_3)\}},
\\\label{37} P&=-\frac{\gamma_3\big(2\mathrm{a}\ddot{\mathrm{a}}+\dot{\mathrm{a}}^2+\mathrm{k}
+\gamma_2\mathrm{a}^2\mathcal{T}_{1}^{1(D)}\big)}{\mathrm{a}^2
\{\gamma_1+\gamma_3(8\pi+5\gamma_1)\}+\gamma_2(\dot{\mathrm{a}}^2+\mathrm{k})+2\gamma_2\mathrm{a}\ddot{\mathrm{a}}}.
\end{align}
There are three unknowns $(\mu,P,\mathrm{a})$ in the above two
equations. Thus we consider a particular functional form (power-law)
of the scale factor, so that we can determine an analytical solution
for the seed source as
\begin{equation}\label{37a}
\mathrm{a}(t)=\mathrm{a}_{0}t^\zeta,
\end{equation}
where $\zeta$ specifies a non-negative constant and $\mathrm{a}_{0}$
refers to the value of the scale factor at present time. Moreover,
the alternate form of anisotropic set \eqref{18}-\eqref{18c} in
terms of FLRW spacetime is attained by using Eq.\eqref{33} as
\begin{align}\label{38}
8\pi\mathcal{Y}^{0}_{0}&=\frac{1}{\mathrm{a}^2}\bigg[\frac{\mathrm{a}
\dot{\mathrm{a}} \dot{\bar{s}}}{\varpi\bar{s}+1}-\bigg(\mathrm{k}
r-\frac{1}{r}\bigg) \bar{s}'-\bigg(3 \mathrm{k}-\frac{1}{r^2}\bigg)
\bar{s}\bigg],\\\label{38a}
8\pi\mathcal{Y}^{1}_{1}&=\frac{\left(1-\mathrm{k} r^2\right)
\bar{s}}{r^2 \mathrm{a}^2},\\\label{38b}
8\pi\mathcal{Y}^{2}_{2}&=\frac{1}{4}\bigg[\frac{6 \dot{\mathrm{a}}
\dot{\bar{s}}}{\mathrm{a}(\varpi\bar{s}+1)}-\frac{2 \left(\mathrm{k}
r^2-1\right) \bar{s}'+4 \mathrm{k} r \bar{s}}{r
\mathrm{a}^2}+\frac{2 \ddot{\bar{s}} (\varpi\bar{s}+1)-3
\varpi\dot{\bar{s}}^2}{(\varpi\bar{s}+1)^2}\bigg],\\\label{38c}
8\pi\mathcal{Y}^{0}_{1}&=\frac{\dot{\bar{s}}}{\varpi r \bar{s}+r}.
\end{align}

We define a linear combination of solutions of both the sources in
such a way that new matter variables delineate an extended
anisotropic solution corresponding to the FLRW metric as
\begin{align}\label{39}
\check{\mu}&=\mu-\varpi\mathcal{Y}^{0}_{0},\\\label{39a}
\check{P}_r&=P+\varpi\mathcal{Y}^{1}_{1},\\\label{39b}
\check{P}_\bot&=P+\varpi\mathcal{Y}^{2}_{2},\\\label{39c}0&=\varpi\mathcal{Y}^{0}_{1},
\end{align}
with anisotropic factor
$\check{\Delta}=\check{P}_\bot-\check{P}_r=\varpi(\mathcal{Y}^{2}_{2}-\mathcal{Y}^{1}_{1})$,
pointing out the disappearance of anisotropy for $\varpi=0$. The
solution corresponding to newly added source is still left to
calculate. We observe that the system \eqref{38}-\eqref{38c} engages
five unknowns
($\mathcal{Y}^{0}_{0},\mathcal{Y}^{1}_{1},\mathcal{Y}^{2}_{2},\mathcal{Y}^{0}_{1},\bar{s}$),
hence another constraints is required to achieve our goal. The
density-like constraint is adopted as the most effective choice in
this scenario as
\begin{equation}\label{40}
\hat{\mu}=\mathcal{Y}^{0}_{0},
\end{equation}
where
$\hat{\mu}=\frac{1}{1+\mu\gamma_2}\big[\big(8\pi+\gamma_1\big)\mu-3P\gamma_1+\gamma_2\mathcal{T}_{0}^{0(D)}\big]$.
The extended variables ($\check{\mu},\check{P}_r,\check{P}_\bot$)
reduce to the isotropic configuration for $\varpi=0$, thus its
non-zero value leads to $\mathcal{Y}^{0}_{1}=0$ which can be
observed from Eq.\eqref{39c}. This ultimately produces
$\dot{\bar{s}}=0$ after using together with Eq.\eqref{38c}. By
substituting Eqs.\eqref{8} and \eqref{38} in \eqref{40}, the
deformation function takes the following form
\begin{equation}\label{41}
\bar{s}=\frac{\dot{\mathrm{a}}^2r^2+\mathrm{k}r^2-1}{(1-\mathrm{k}r^2)}
+\frac{C_0}{r(1-\mathrm{k}r^2)},
\end{equation}
where $C_0$ refers to an integration constant. The ultimate
expressions for the state determinants \eqref{39}-\eqref{39b} and
anisotropy are
\begin{align}\nonumber
\check{\mu}&=\frac{\gamma_2\mathrm{a}^2\mathcal{T}_{0}^{0(D)}-3\left(\dot{\mathrm{a}}^2+\mathrm{k}\right)}
{3\gamma_2(\dot{\mathrm{a}}^2+\mathrm{k})-\mathrm{a}^2\{8\pi+\gamma_1(1-3\gamma_3)\}}+\frac{\varpi}{8\pi\mathrm{a}^2}\\\label{42}
&\times\bigg[\frac{1}{r^2}-3\mathrm{k}-\dot{\mathrm{a}}^2\bigg\{3+\frac{2r^3\mathrm{a}\ddot{\mathrm{a}}}{r(\mathrm{k}r^2-1)
(\varpi-1)+C_0\varpi+r^3\varpi\dot{\mathrm{a}}^2}\bigg\}\bigg],\\\nonumber
\check{P}_r&=-\frac{\gamma_3\big\{2\mathrm{a}\ddot{\mathrm{a}}+\dot{\mathrm{a}}^2+\mathrm{k}
+\gamma_2\mathrm{a}^2\mathcal{T}_{1}^{1(D)}\big\}}{\mathrm{a}^2
\{\gamma_1+\gamma_3(8\pi+5\gamma_1)\}+\gamma_2(\dot{\mathrm{a}}^2+\mathrm{k})+2\gamma_2\mathrm{a}\ddot{\mathrm{a}}}\\\label{42a}
&+\frac{\varpi\big\{C_0-r+\mathrm{k}r^3+r^3\dot{\mathrm{a}}^2\big\}}{8\pi\mathrm{a}^2r^3},\\\nonumber
\check{P}_\bot&=-\frac{\gamma_3\big\{2\mathrm{a}\ddot{\mathrm{a}}+\dot{\mathrm{a}}^2+\mathrm{k}
+\gamma_2\mathrm{a}^2\mathcal{T}_{1}^{1(D)}\big\}}{\mathrm{a}^2
\{\gamma_1+\gamma_3(8\pi+5\gamma_1)\}+\gamma_2(\dot{\mathrm{a}}^2+\mathrm{k})+2\gamma_2\mathrm{a}\ddot{\mathrm{a}}}
+\frac{\varpi}{32\pi}\\\nonumber
&\times\bigg[\frac{4\mathrm{k}r^3-2C_0+4r^3\dot{\mathrm{a}}^2}{\mathrm{a}^2r^3}
+\frac{12r^3\dot{\mathrm{a}}^2\ddot{\mathrm{a}}}{\mathrm{a}\big\{r(\mathrm{k}r^2-1)(\varpi-1)+C_0\varpi
+r^3\varpi\dot{\mathrm{a}}^2\big\}}\\\nonumber
&+\frac{4r^3}{\big\{r(\mathrm{k}r^2-1)(\varpi-1)+C_0\varpi
+r^3\varpi\dot{\mathrm{a}}^2\big\}^2}\big\{\big(r(\mathrm{k}r^2-1)(\varpi-1)\\\label{42b}
&+C_0\varpi-2r^3\varpi\dot{\mathrm{a}}^2\big)\ddot{\mathrm{a}}^2+\dot{\mathrm{a}}\dddot{\mathrm{a}}
\big(r(\mathrm{k}r^2-1)(\varpi-1)+C_0\varpi
+r^3\varpi\dot{\mathrm{a}}^2\big)\big\}\bigg],\\\nonumber
\check{\Delta}&=\frac{\varpi}{32\pi{r^3}}\bigg[\frac{12r^6\dot{\mathrm{a}}^2\ddot{\mathrm{a}}}{\mathrm{a}\big\{r(\mathrm{k}r^2-1)
(\varpi-1)+C_0\varpi+r^3\varpi\dot{\mathrm{a}}^2\big\}}+\frac{4r-6C_0}{\mathrm{a}^2}\\\nonumber
&+\frac{4r^6}{\big\{r(\mathrm{k}r^2-1)(\varpi-1)+C_0\varpi
+r^3\varpi\dot{\mathrm{a}}^2\big\}^2}\big\{\big(r(\mathrm{k}r^2-1)(\varpi-1)\\\label{42c}
&+C_0\varpi-2r^3\varpi\dot{\mathrm{a}}^2\big)\ddot{\mathrm{a}}^2+\dot{\mathrm{a}}\dddot{\mathrm{a}}
\big(r(\mathrm{k}r^2-1)(\varpi-1)+C_0\varpi
+r^3\varpi\dot{\mathrm{a}}^2\big)\big\}\bigg].
\end{align}
\begin{figure}\center
\epsfig{file=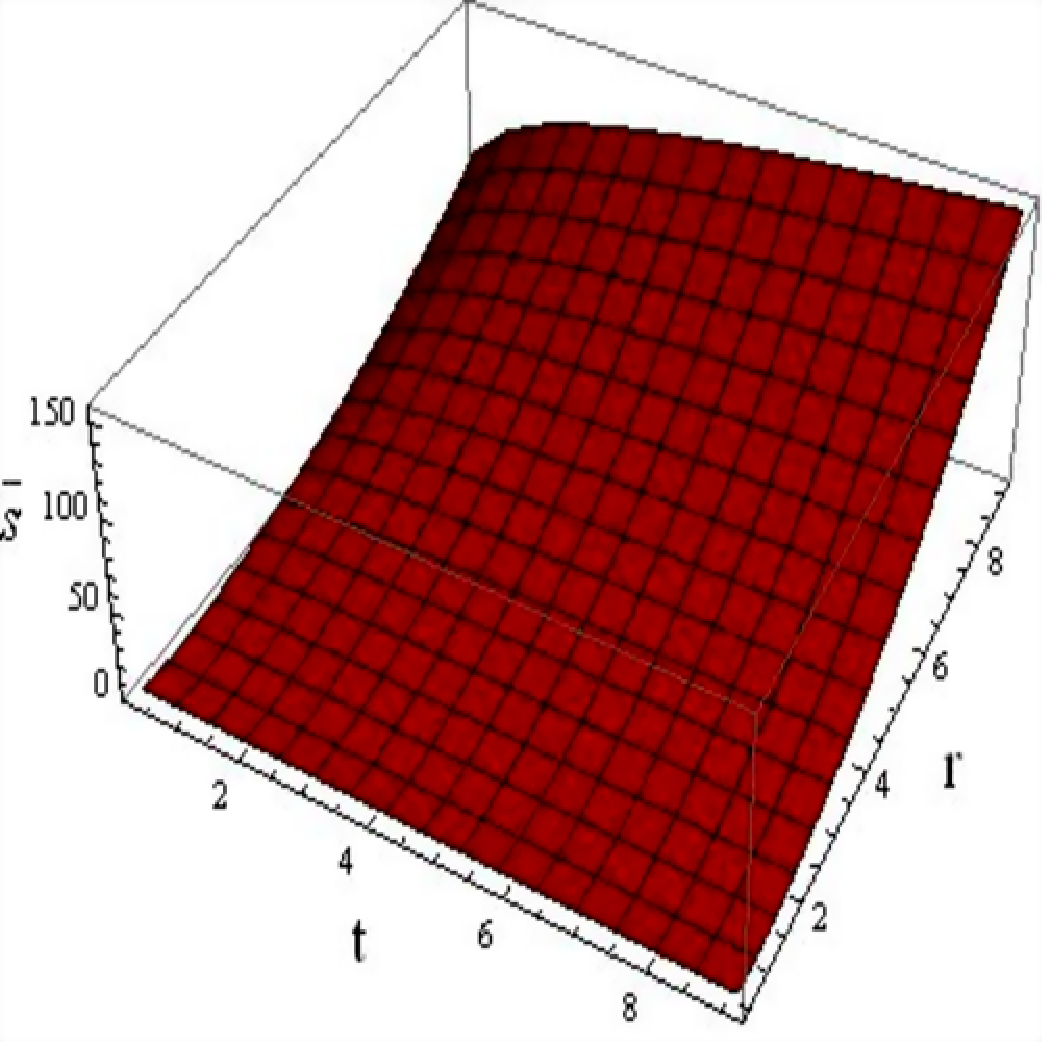,width=0.4\linewidth}
\caption{Plot of deformation function.}
\end{figure}

The physical attributes corresponding to the extended FLRW solution
\eqref{42}-\eqref{42c} are now investigated for multiple values of
the parameter $\gamma_3$. Recent observations disclose the
deceleration parameter within the interval $[-1,0]$, which
ultimately leads to $\zeta>1$ \cite{24ab}, therefore we choose it
$1.1$ together with $C_0=2$ and $\mathrm{k}=0$. Furthermore, we
explore the impact of the decoupling parameter and modified gravity
on different models of the universe by taking
$\gamma_1=0.05,~\gamma_2=0.01$ and $\varpi=0.01,~0.05$. Figure
\textbf{1} manifests the plot of deformation function \eqref{41}
which vanishes initially and then increases with the rise in $t$ and
$r$.

The energy conditions $(\mathbb{EC}s)$ are the constraints on state
determinants (defined by $\mathbb{EMT}$) whose fulfilment (or
dissatisfaction) ensures the presence of usual (or exotic) matter
inside a compact structure. We analyze graphical behavior of such
bounds to guarantee the viability of our developed model. These
conditions are generally classified as
\begin{align}\nonumber
\text{Weak}~\mathbb{EC}s:\quad&\check{\mu}\geq0,\quad\check{\mu}+\check{P}_\bot\geq0,\quad\check{\mu}+\check{P_r}\geq0,\\\nonumber
\text{Null}~\mathbb{EC}s:\quad&\check{\mu}+\check{P}_\bot\geq0,\quad\check{\mu}+\check{P_r}\geq0,\\\nonumber
\text{Strong}~\mathbb{EC}:\quad&\check{\mu}+\check{P_r}+2\check{P}_\bot\geq0,\\\label{59}
\text{Dominant}~\mathbb{EC}s:\quad&\check{\mu}-\check{P_r}\geq0,\quad\check{\mu}-\check{P}_\bot\geq0.
\end{align}

Further, we employ different approaches to discuss stable regions of
the resulting solutions. They are defined as follows
\begin{itemize}
\item Causality Condition: It provides the stable system only if $v_{s}^2\in(0,1)$, i.e., speed of sound should always be less
than the speed of light in a medium \cite{24a}.
\item Herrera Cracking Approach: It ensures the stability of a celestial object only when an inequality
$0<|v^{2}_{s\bot}-v^{2}_{sr}|<1$ satisfies, where
$v^2_{sr}=\frac{d\check{P}_{r}}{d\check{\mu}}$ is radial and
$v^2_{s\bot}=\frac{d\check{P}_{\bot}}{d\check{\mu}}$ refers to the
tangential sound speed \cite{24}.
\end{itemize}

The following subsections explore the behavior of the developed
solution corresponding to three different eras.

\subsection{Radiation-Dominated Phase}

The matter formed by several relativistic particles, namely
neutrinos and photons, etc. dominated our universe in this era. The
$\mathbb{E}o\mathbb{S}$ \eqref{16a} helps to discuss this epoch for
the parametric value as $\gamma_3=\frac{1}{3}$, which discloses that
the entire energy density at that time was equivalent to three times
the pressure. The ratio of mass to the momentum of the fluid was
also observed to be less than $1$ at that time. The nature of
physical variables and anisotropy is displayed in Figure \textbf{2}
corresponding to both values of $\varpi$. The decreasing behavior of
the energy density with time suggests an expanding universe (upper
left plot). Also, we observe an increasing profile of density and
decreasing pressure components with the increment in $\varpi$. The
last plot of Figure \textbf{2} exhibits that the pressure anisotropy
is negative at some initial time, and then shows a positive trend at
a later time. Figure \textbf{3} displays the positive behavior of
all $\mathbb{EC}s$, hence representing a viable model. All the
criteria to check stability are plotted in Figure \textbf{4} whose
fulfillment portrays the extended model to be stable everywhere for
$\gamma_3=\frac{1}{3}$.
\begin{figure}\center
\epsfig{file=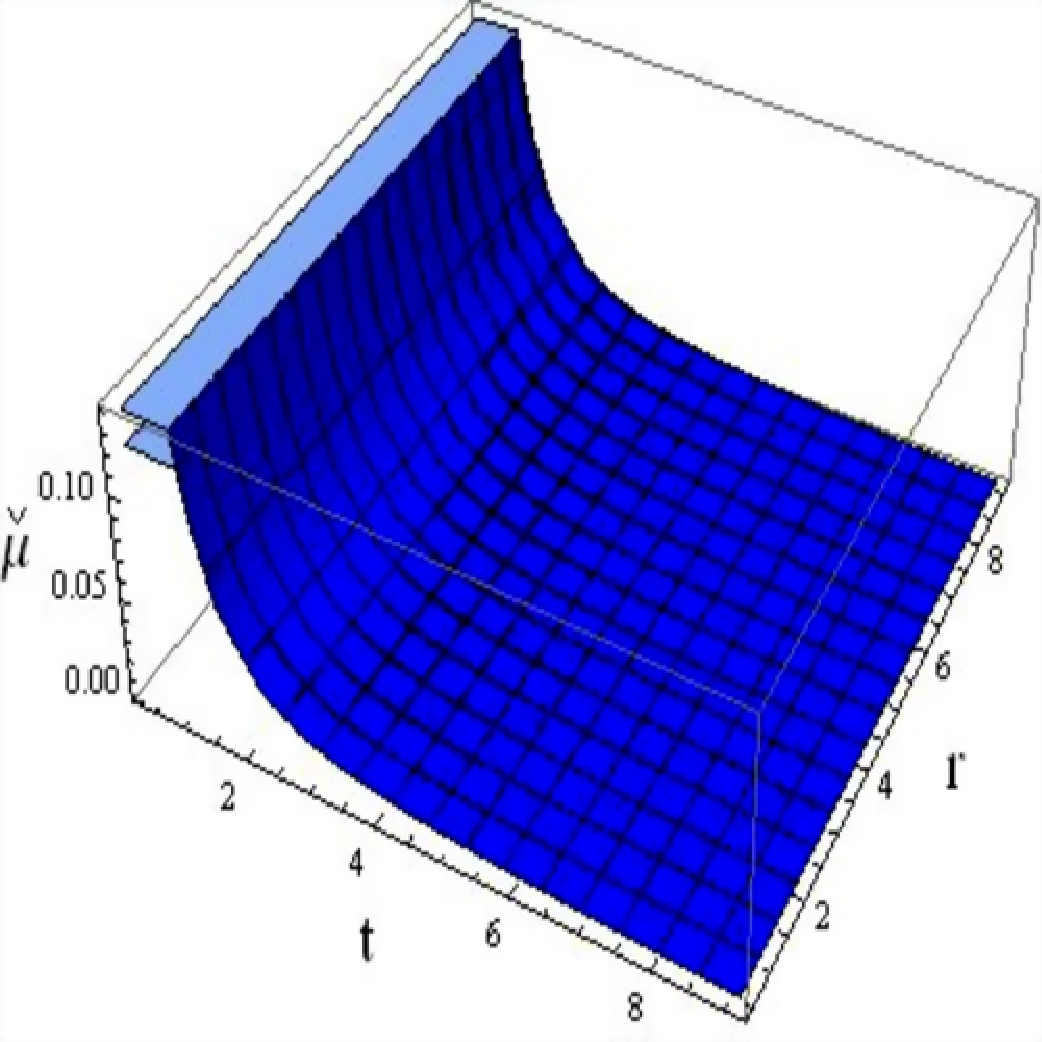,width=0.4\linewidth}\epsfig{file=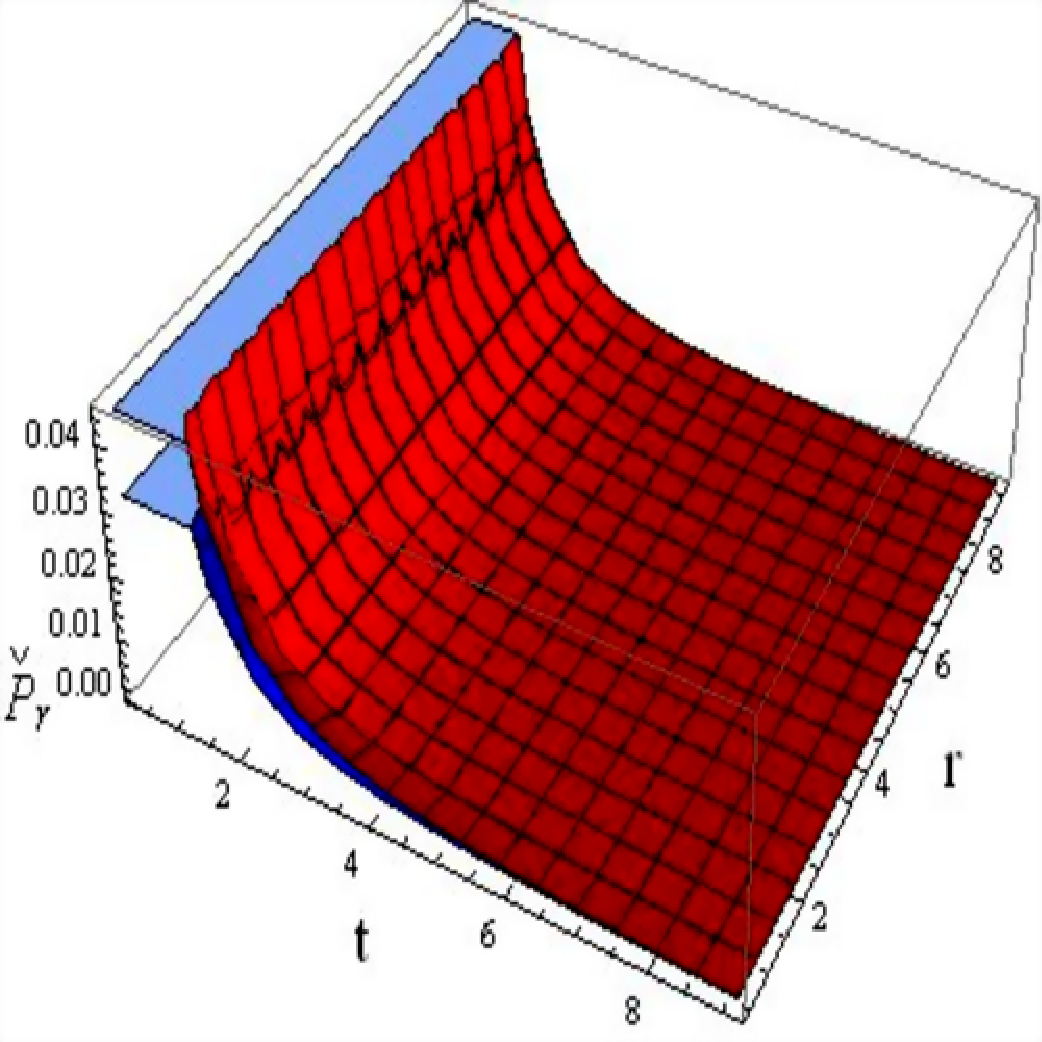,width=0.4\linewidth}
\epsfig{file=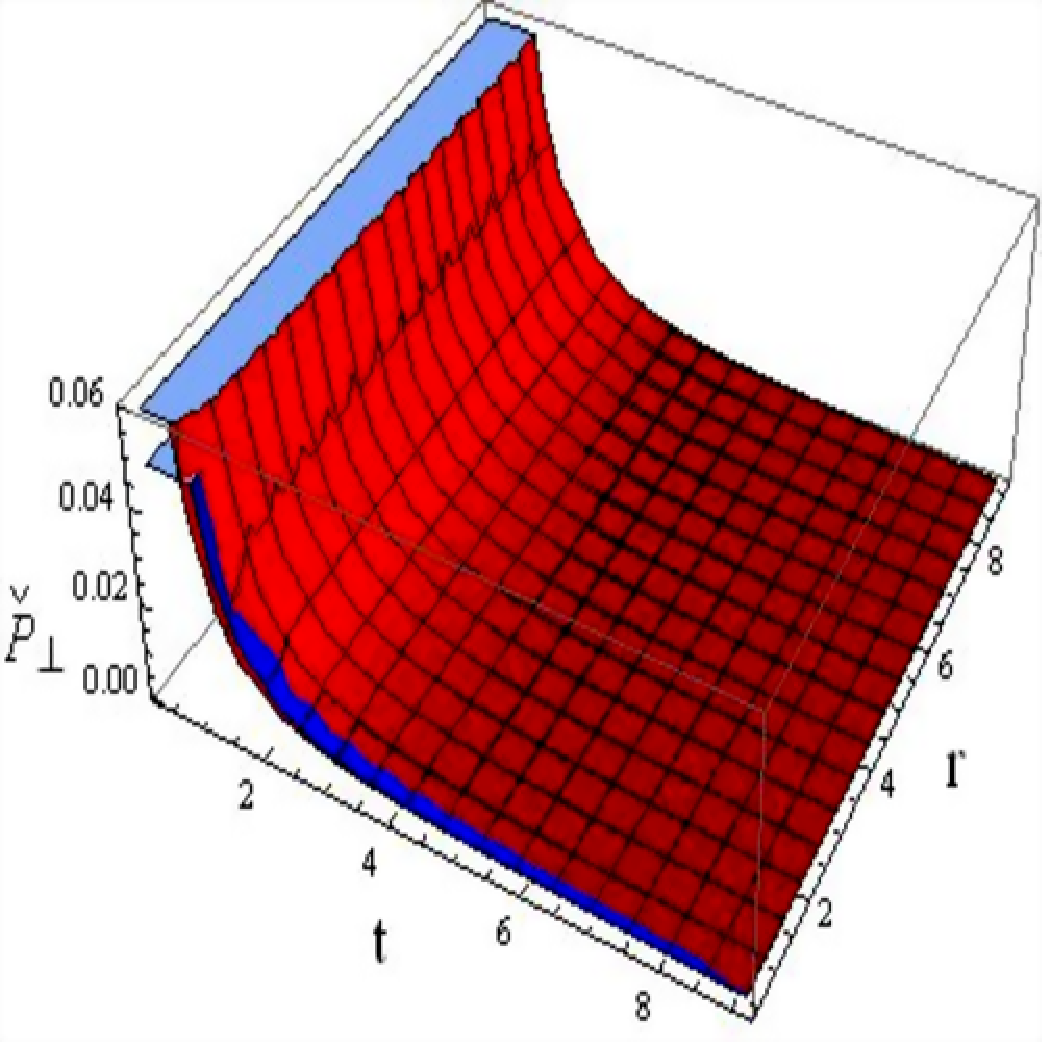,width=0.4\linewidth}\epsfig{file=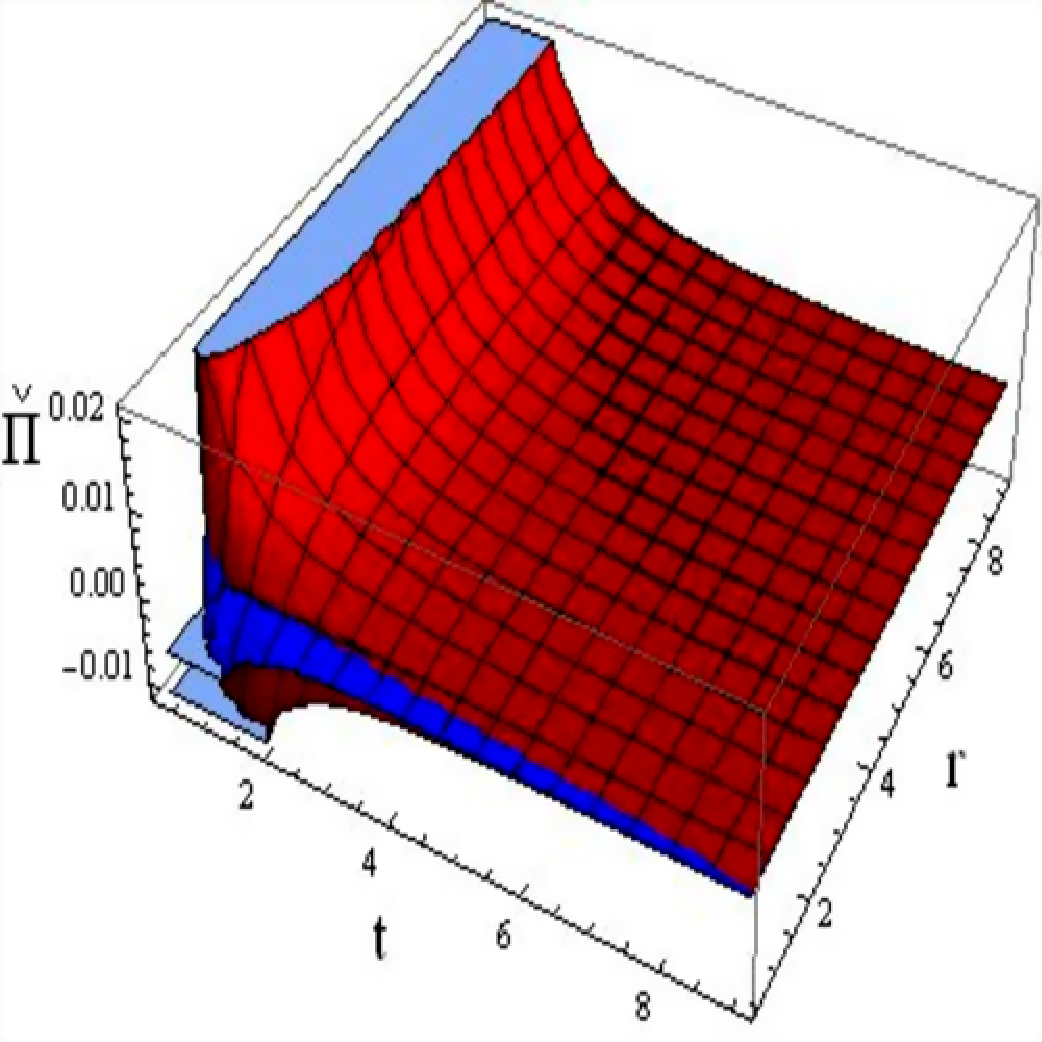,width=0.4\linewidth}
\caption{Plots of matter determinants and anisotropy (in km$^{-2}$)
with $\gamma_3=\frac{1}{3}$ for $\varpi=0.1$ (blue) and $0.7$
(red).}
\end{figure}
\begin{figure}\center
\epsfig{file=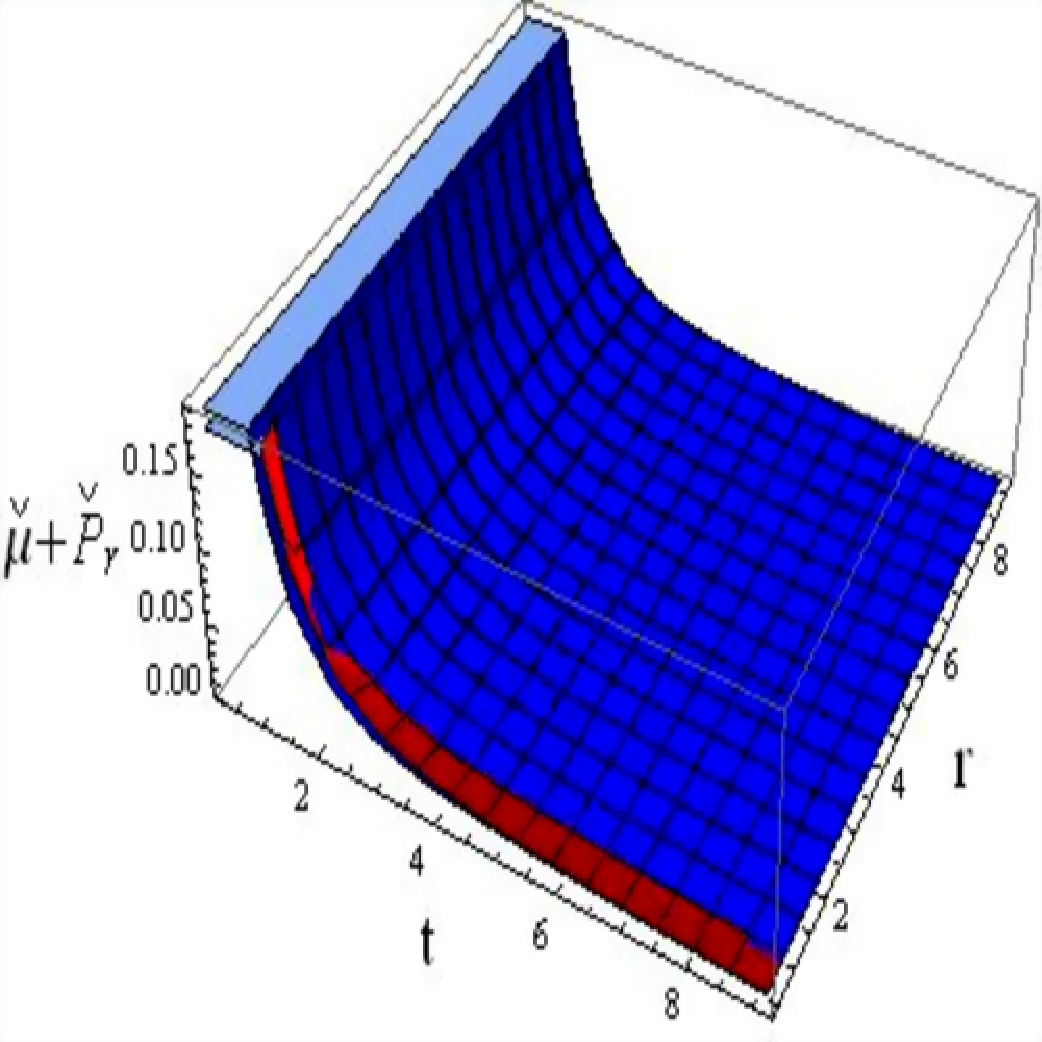,width=0.4\linewidth}\epsfig{file=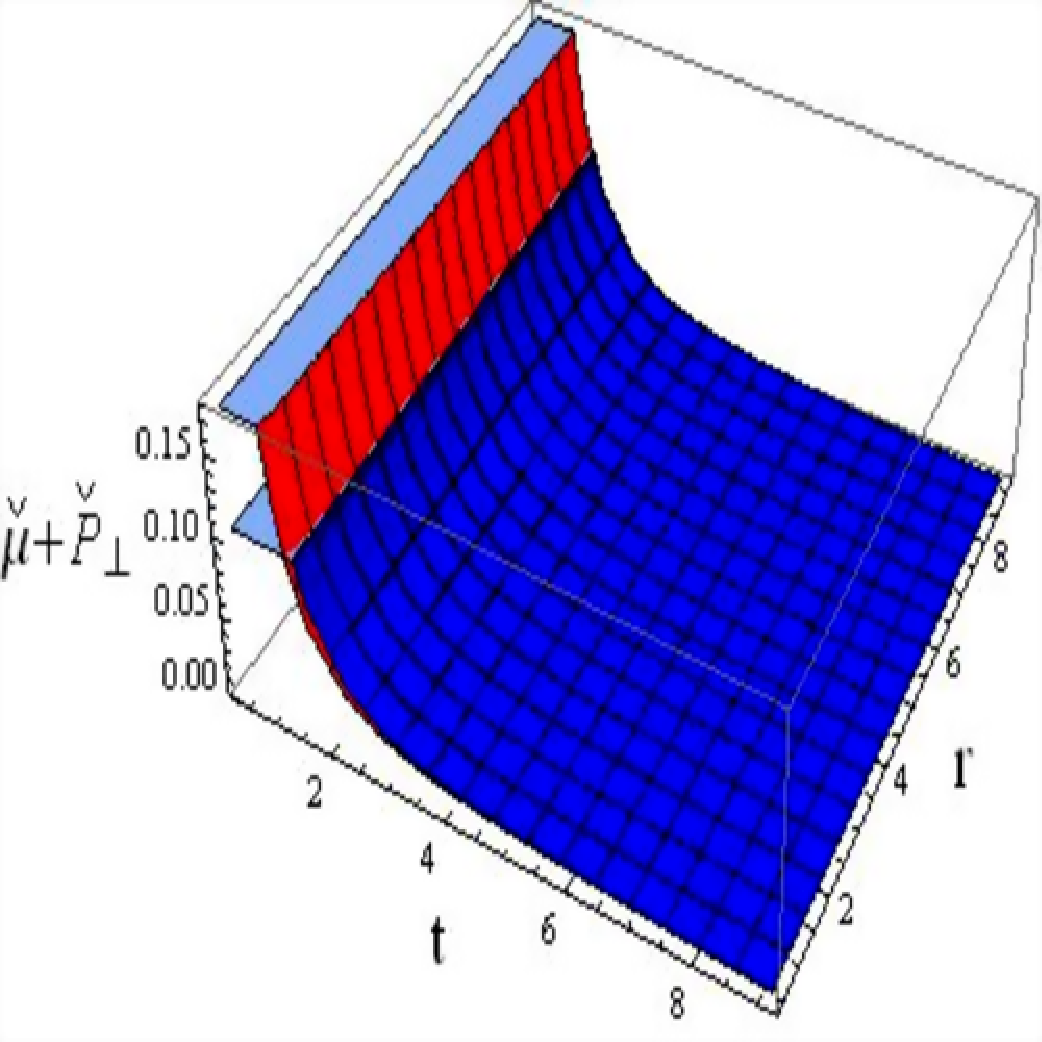,width=0.4\linewidth}
\epsfig{file=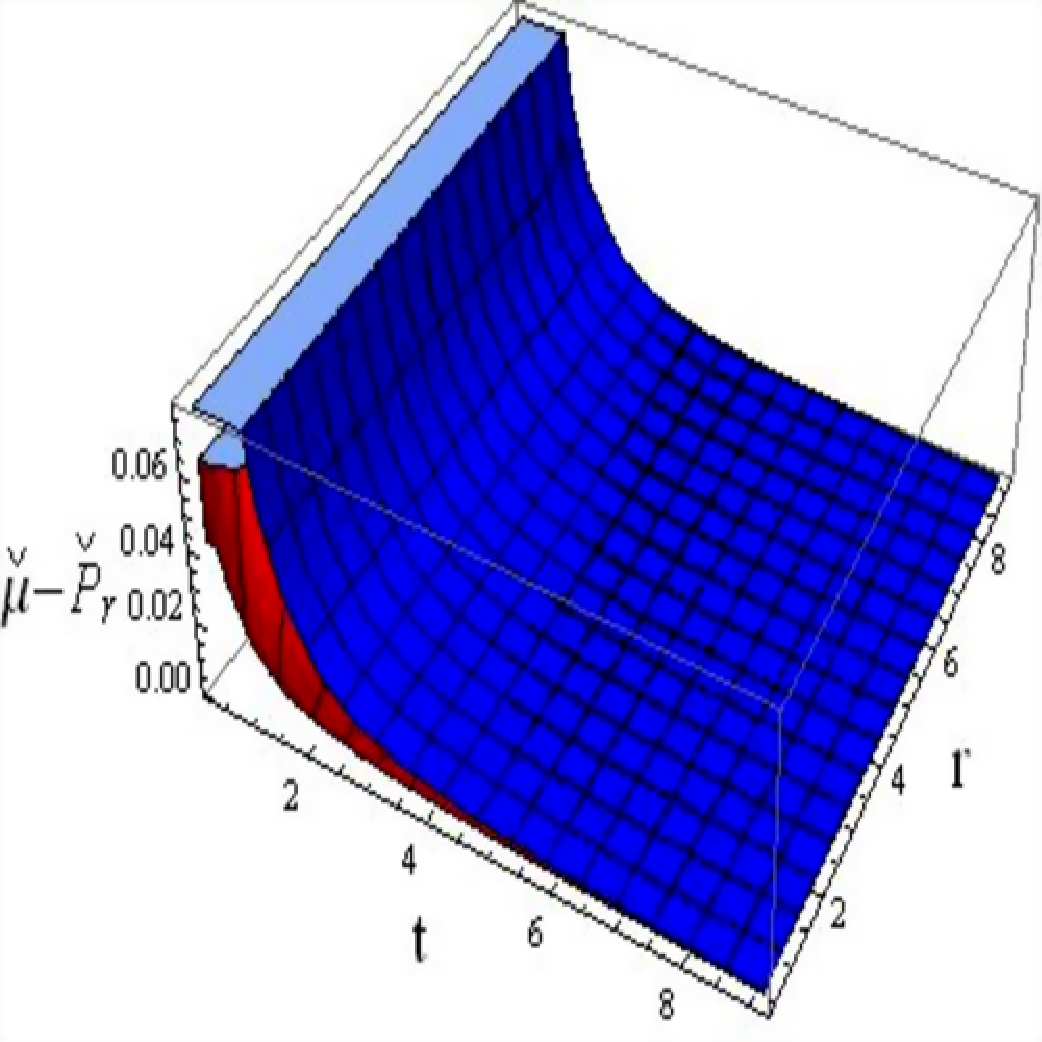,width=0.4\linewidth}\epsfig{file=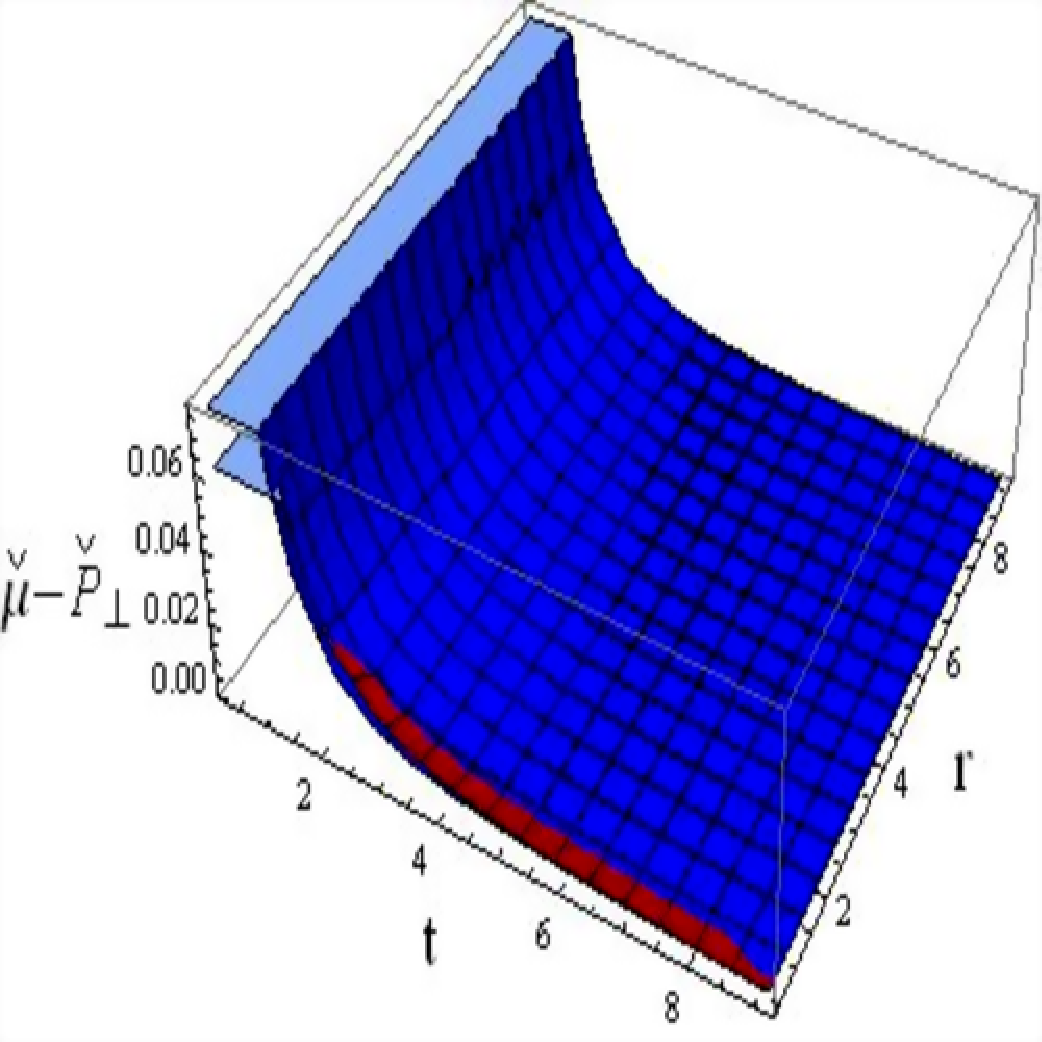,width=0.4\linewidth}
\epsfig{file=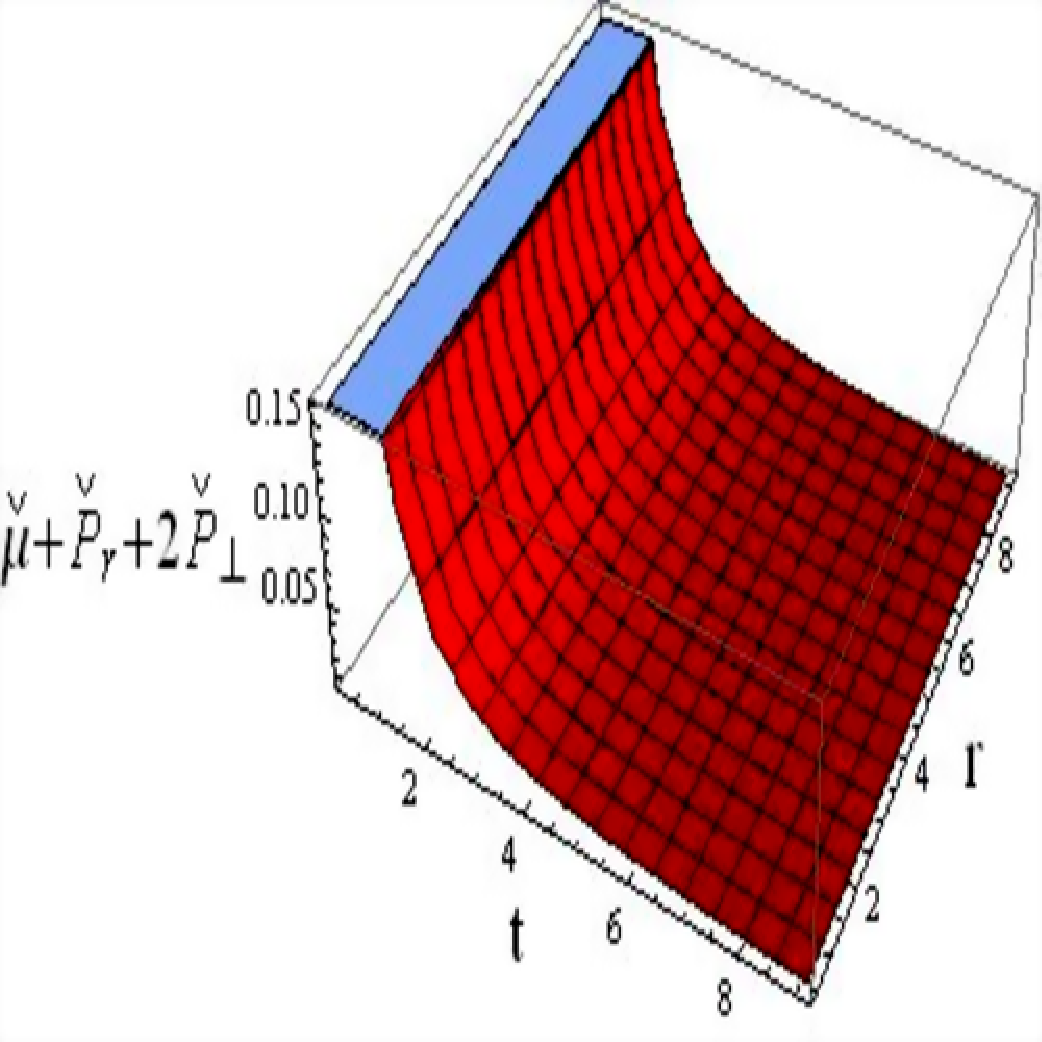,width=0.4\linewidth} \caption{Plots of
$\mathbb{EC}s$ (in km$^{-2}$) with $\gamma_3=\frac{1}{3}$ for
$\varpi=0.1$ (blue) and $0.7$ (red).}
\end{figure}
\begin{figure}\center
\epsfig{file=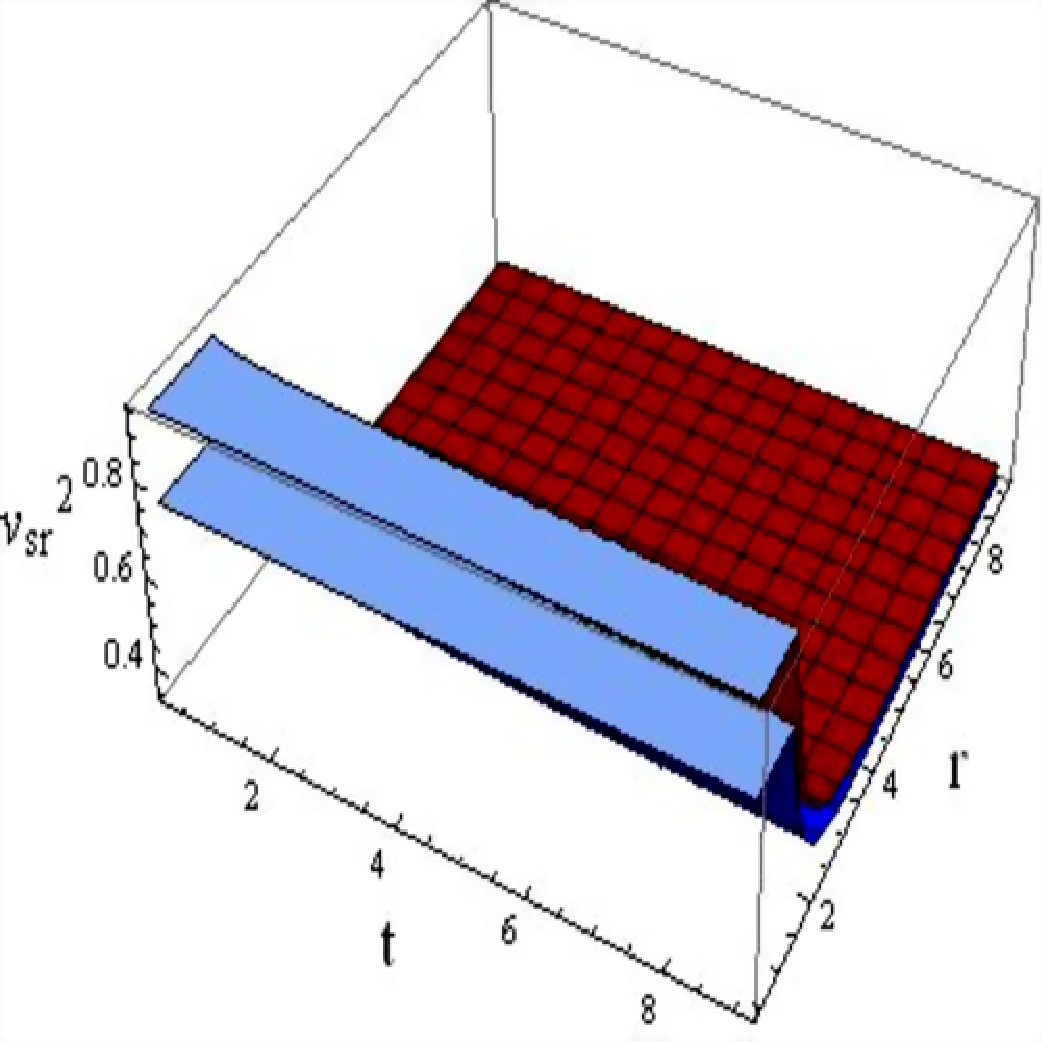,width=0.4\linewidth}\epsfig{file=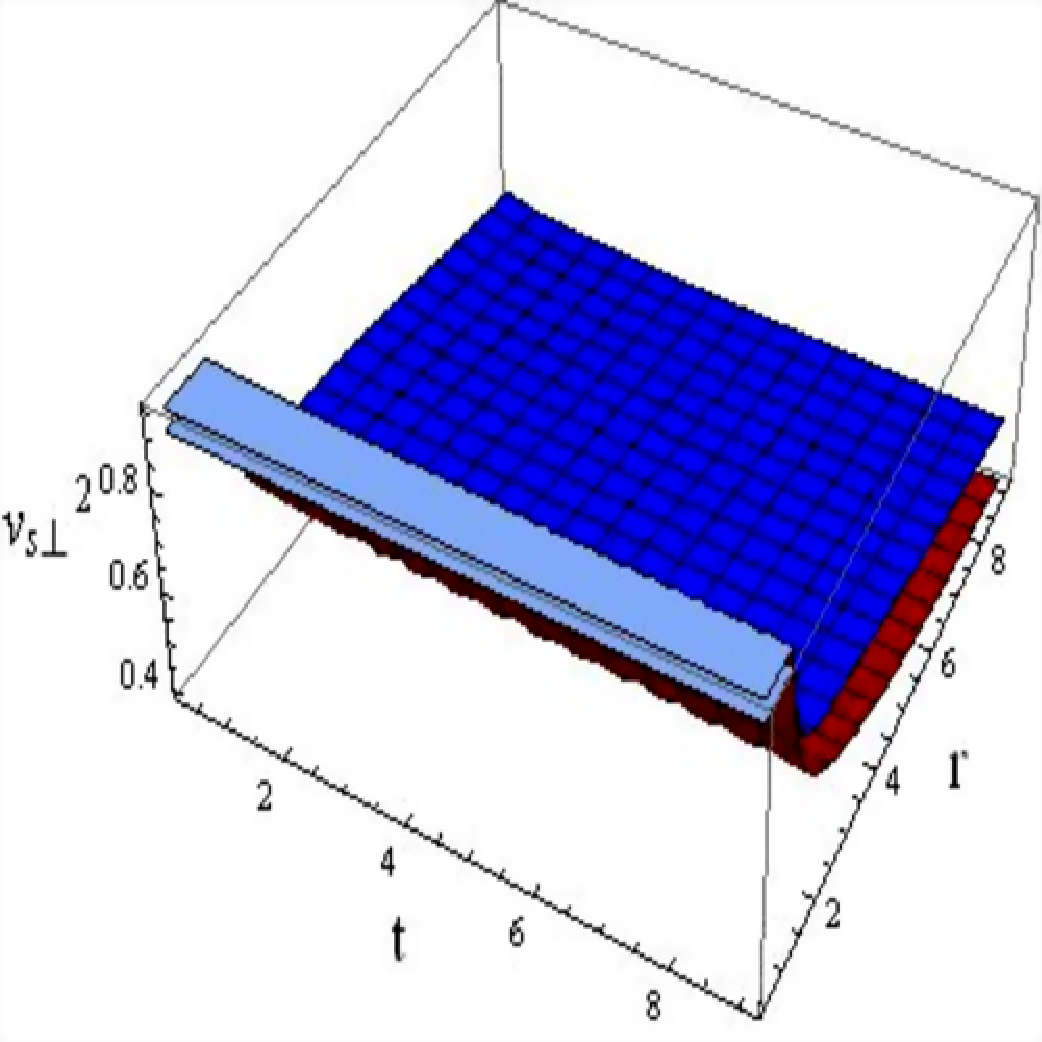,width=0.4\linewidth}
\epsfig{file=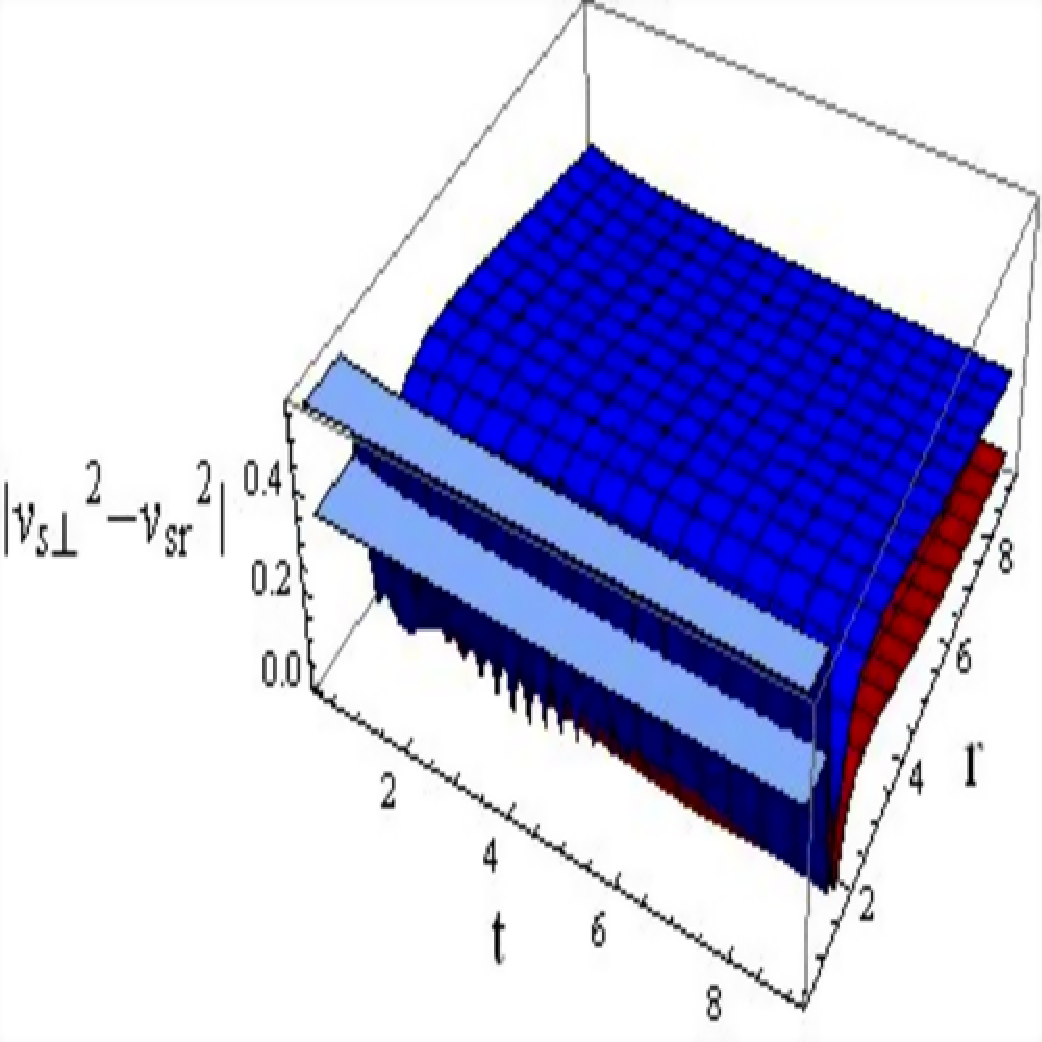,width=0.4\linewidth} \caption{Plots of
$v^{2}_{sr}$, $v^{2}_{s\bot}$ and $|v^{2}_{s\bot}-v^{2}_{sr}|$ with
$\gamma_3=\frac{1}{3}$ for $\varpi=0.1$ (blue) and $0.7$ (red).}
\end{figure}

\subsection{Matter-Dominated Phase}

The non-relativistic particles, such as baryons whose kinetic energy
is much lesser than their mass energy, dominated a considerable
portion of the universe during this era. These elementary particles
merge to make a substance, and we define it as dust. The physical
quantities and anisotropic factor are graphically interpreted in
Figure \textbf{5}. The universe's expansion in this era can also be
observed as the energy density exhibits decreasing behavior with the
increase in time and is directly related to $\varpi$ (upper left).
Since Eq.\eqref{16a} with $\gamma_3=0$ leads to this particular
phase, therefore the pressure component is observed to be
negligible. Nevertheless, both radial and tangential pressures show
a positive trend and then disappear over time. The $\mathbb{EC}s$
are plotted in Figure \textbf{6} that represent acceptable behavior,
leading to a viable model. The plot of tangential sound speed
expresses that the corresponding solution is unstable near the
center for both values of $\varpi$, whereas Herrera cracking
approach exhibits the same behavior only for $\varpi=0.01$ (Figure
\textbf{7}).
\begin{figure}\center
\epsfig{file=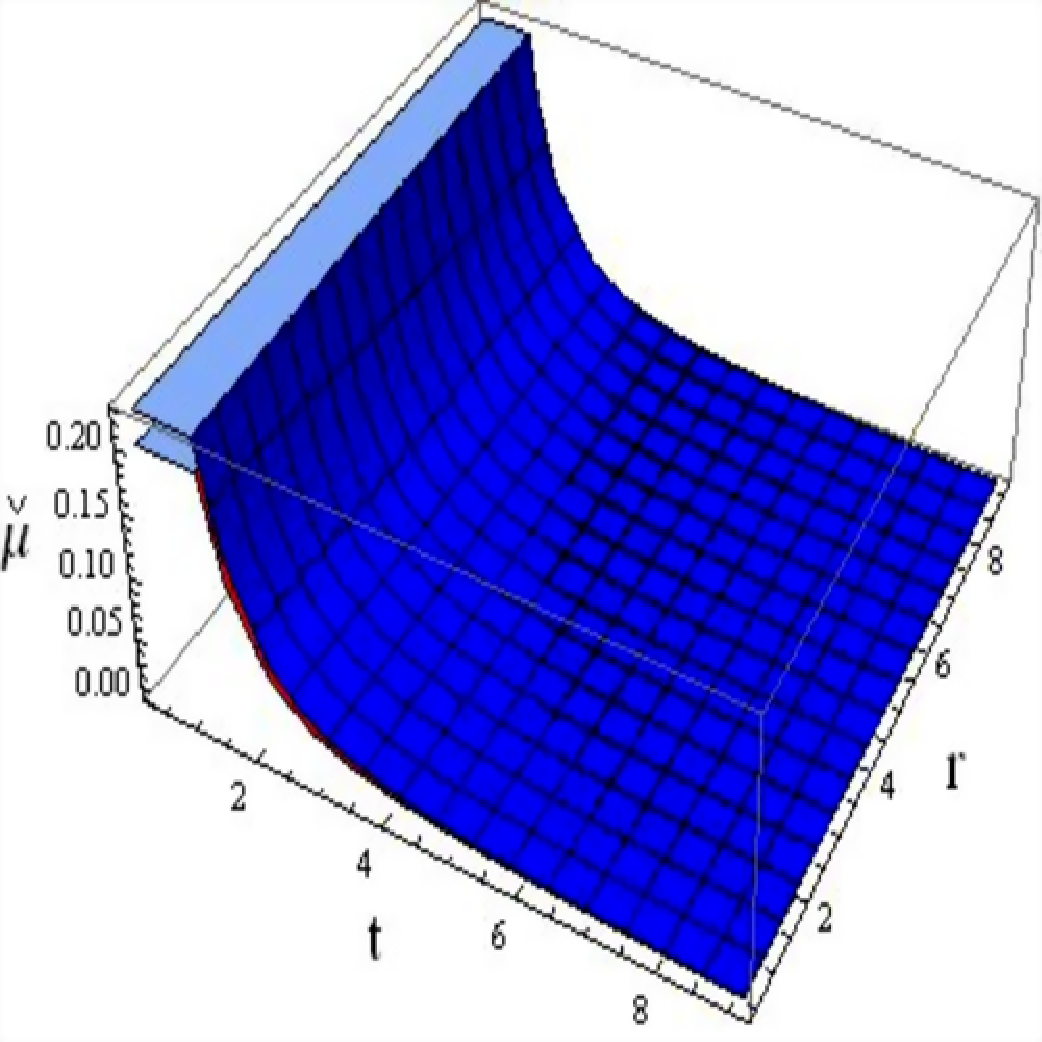,width=0.4\linewidth}\epsfig{file=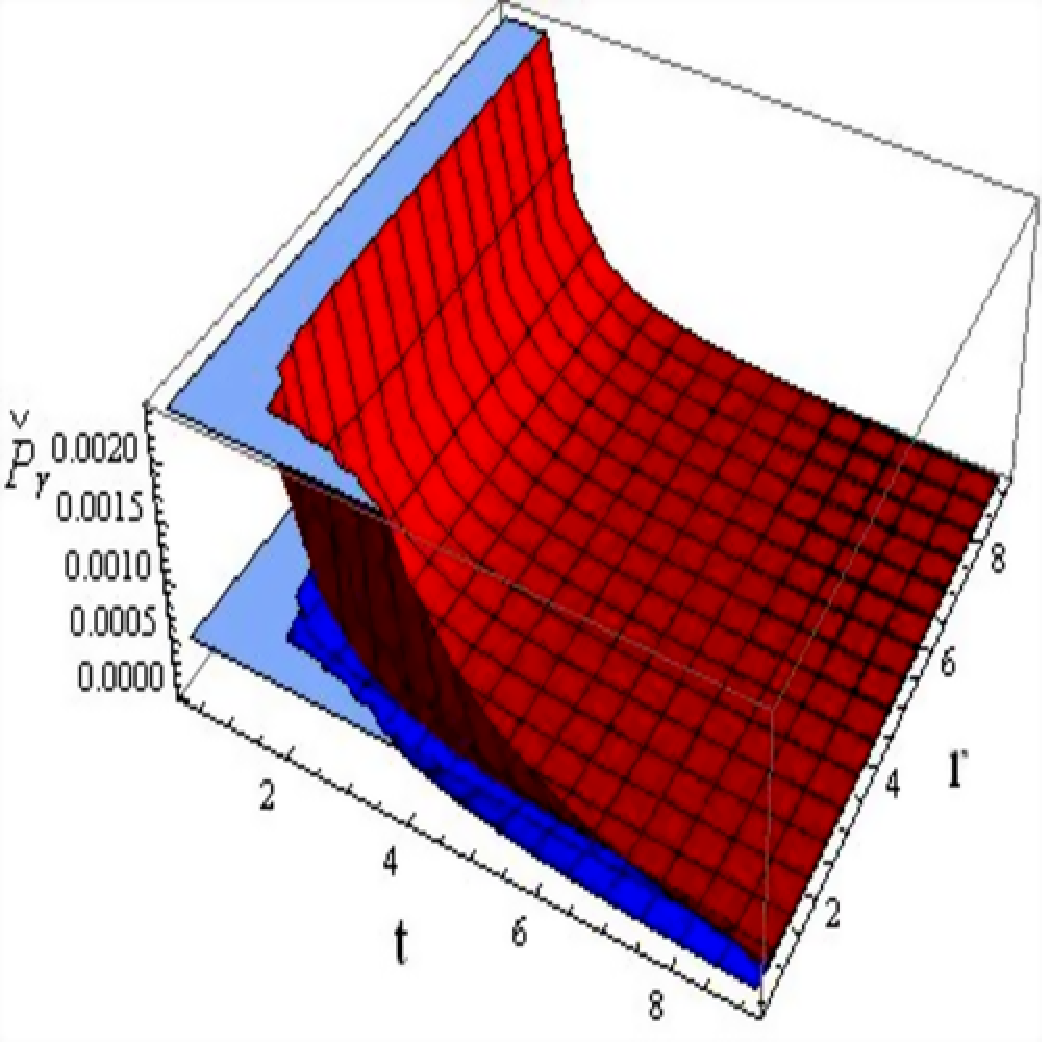,width=0.4\linewidth}
\epsfig{file=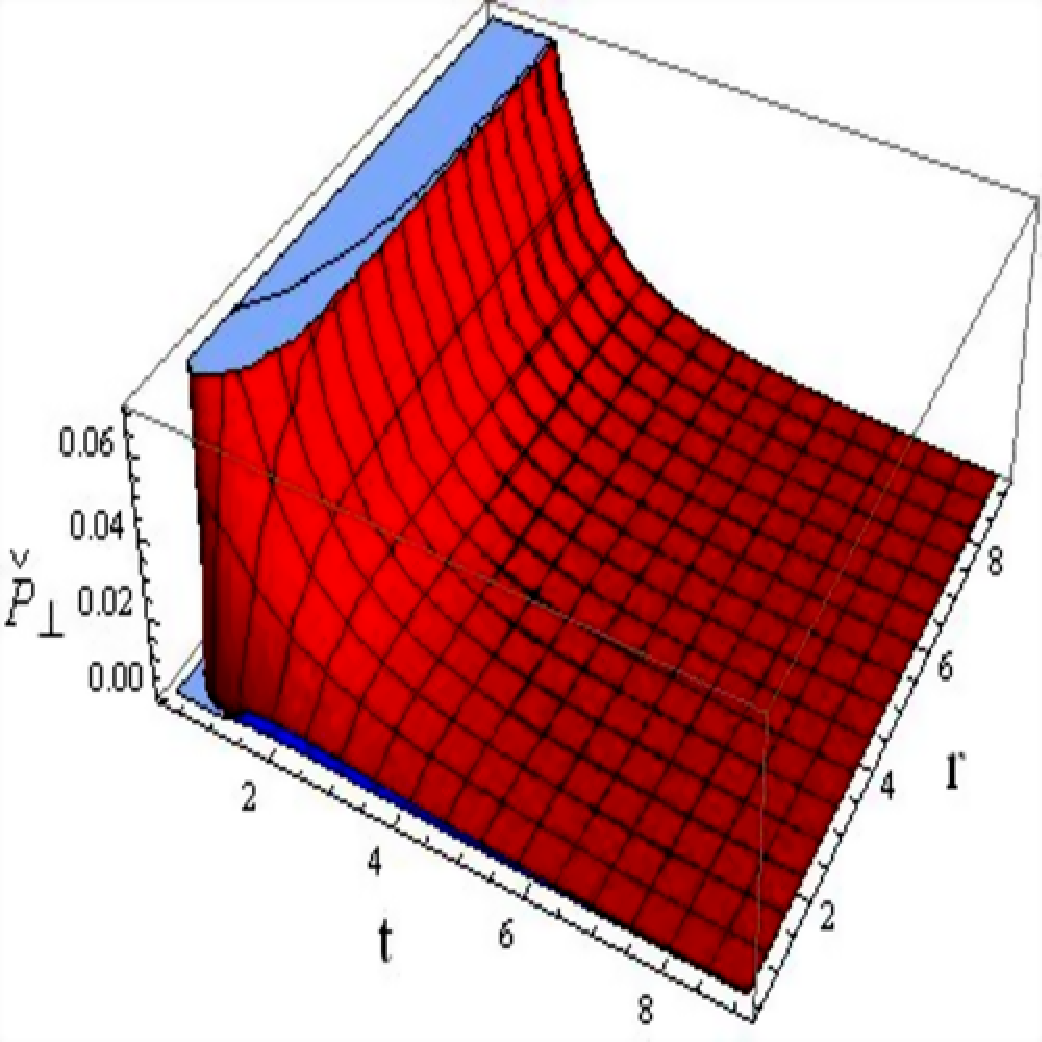,width=0.4\linewidth}\epsfig{file=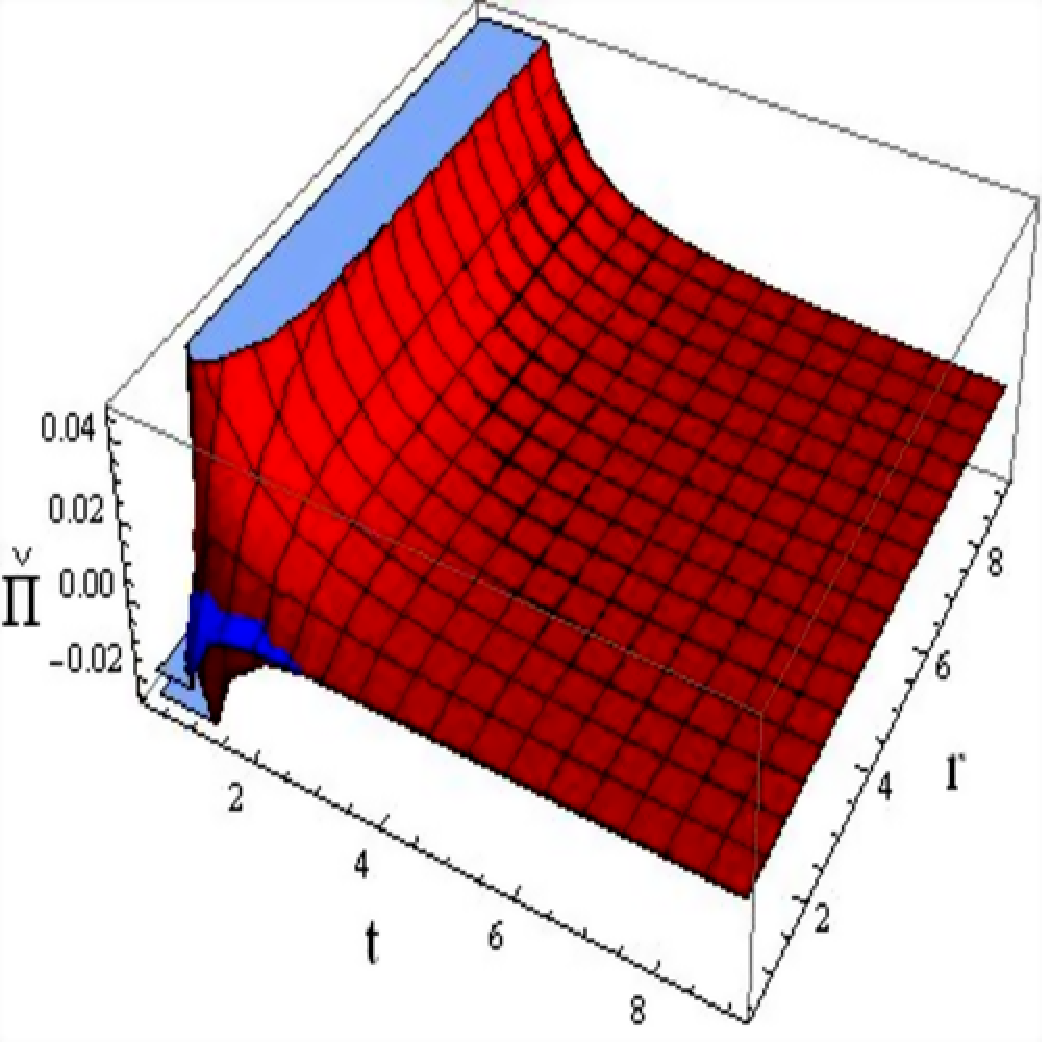,width=0.4\linewidth}
\caption{Plots of matter determinants and anisotropy (in km$^{-2}$)
with $\gamma_3=0$ for $\varpi=0.1$ (blue) and $0.7$ (red).}
\end{figure}
\begin{figure}\center
\epsfig{file=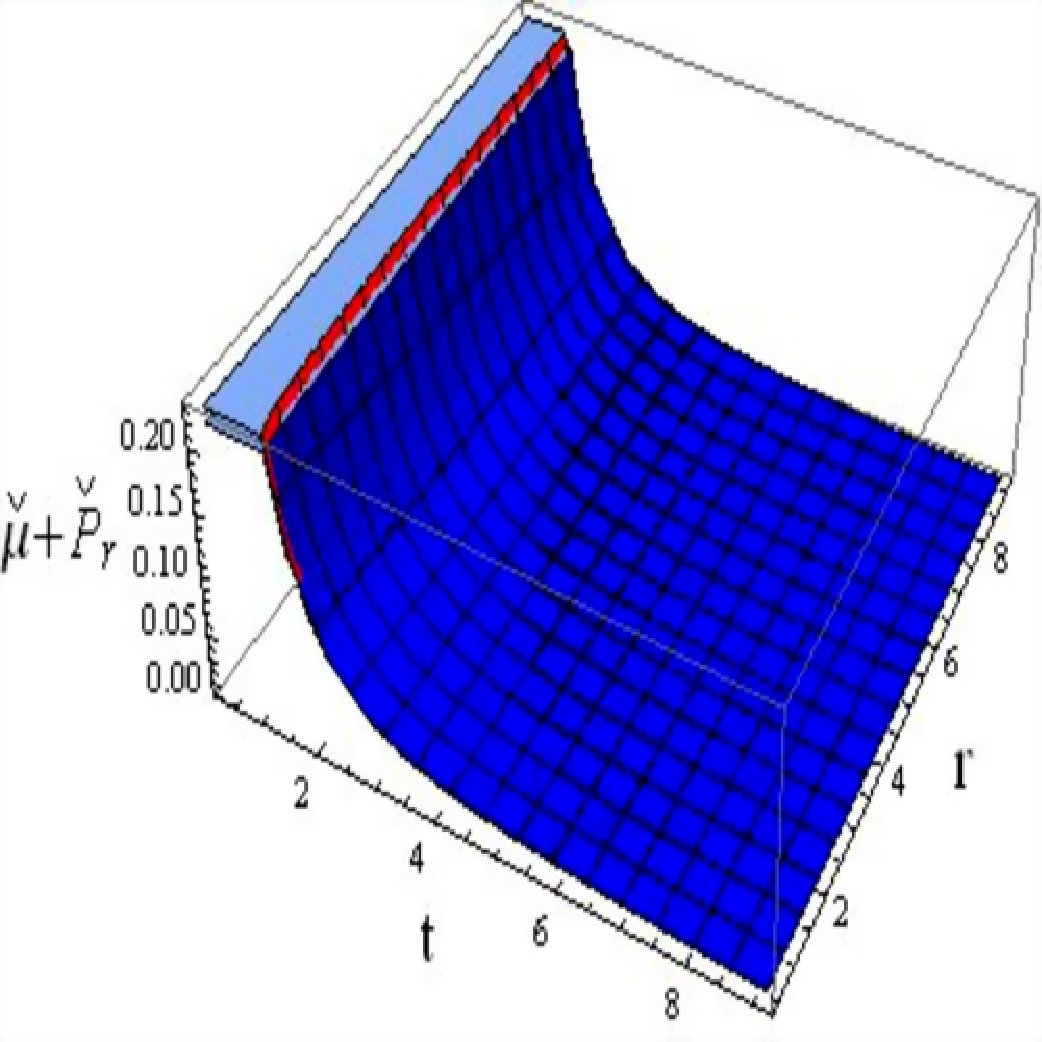,width=0.4\linewidth}\epsfig{file=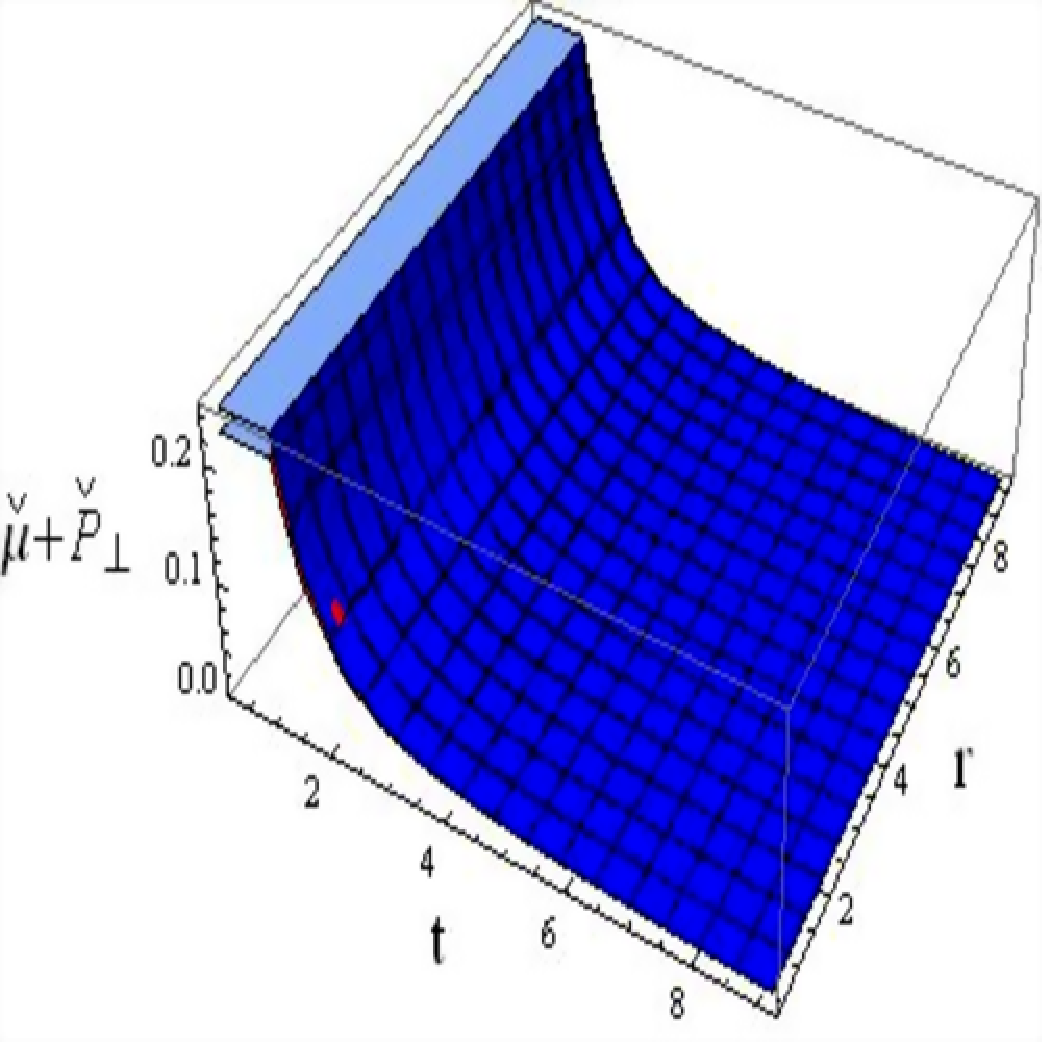,width=0.4\linewidth}
\epsfig{file=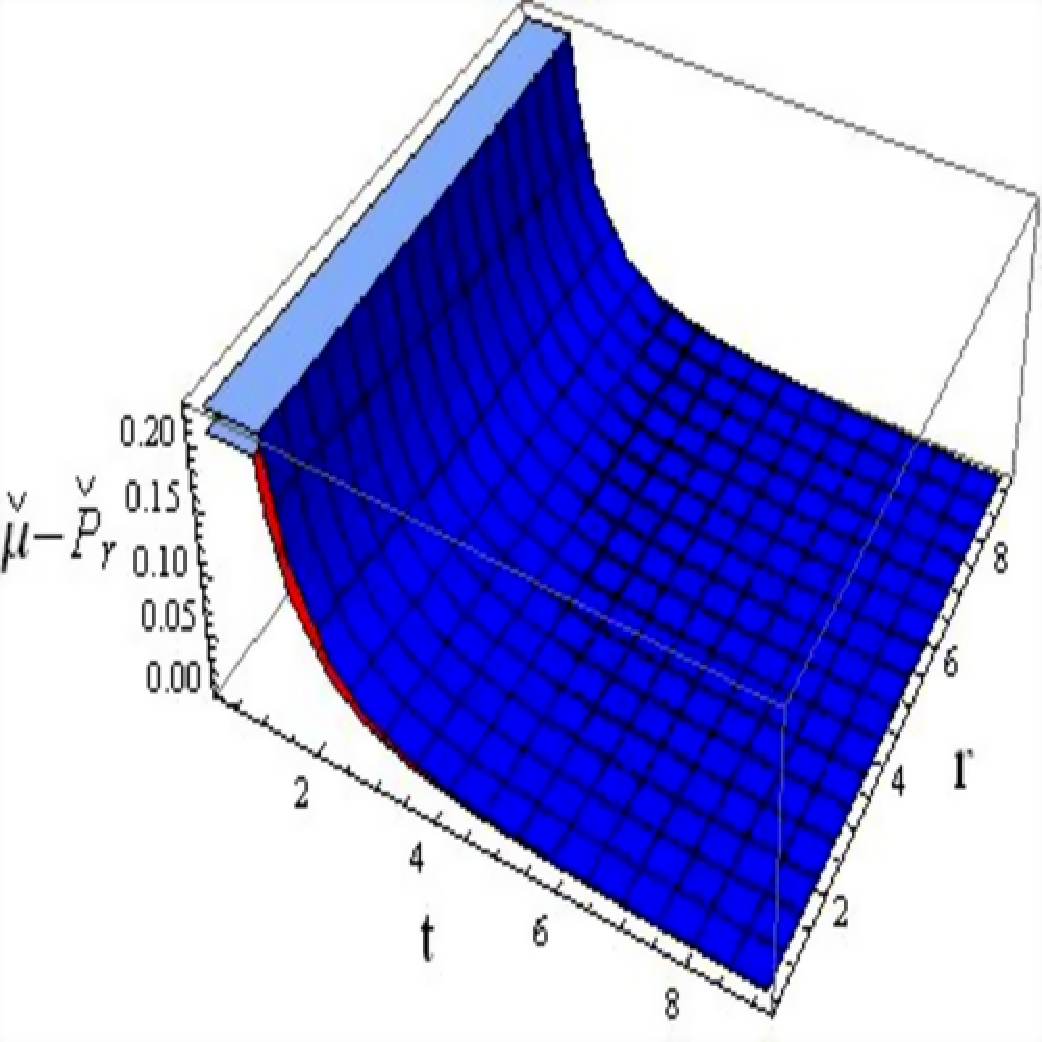,width=0.4\linewidth}\epsfig{file=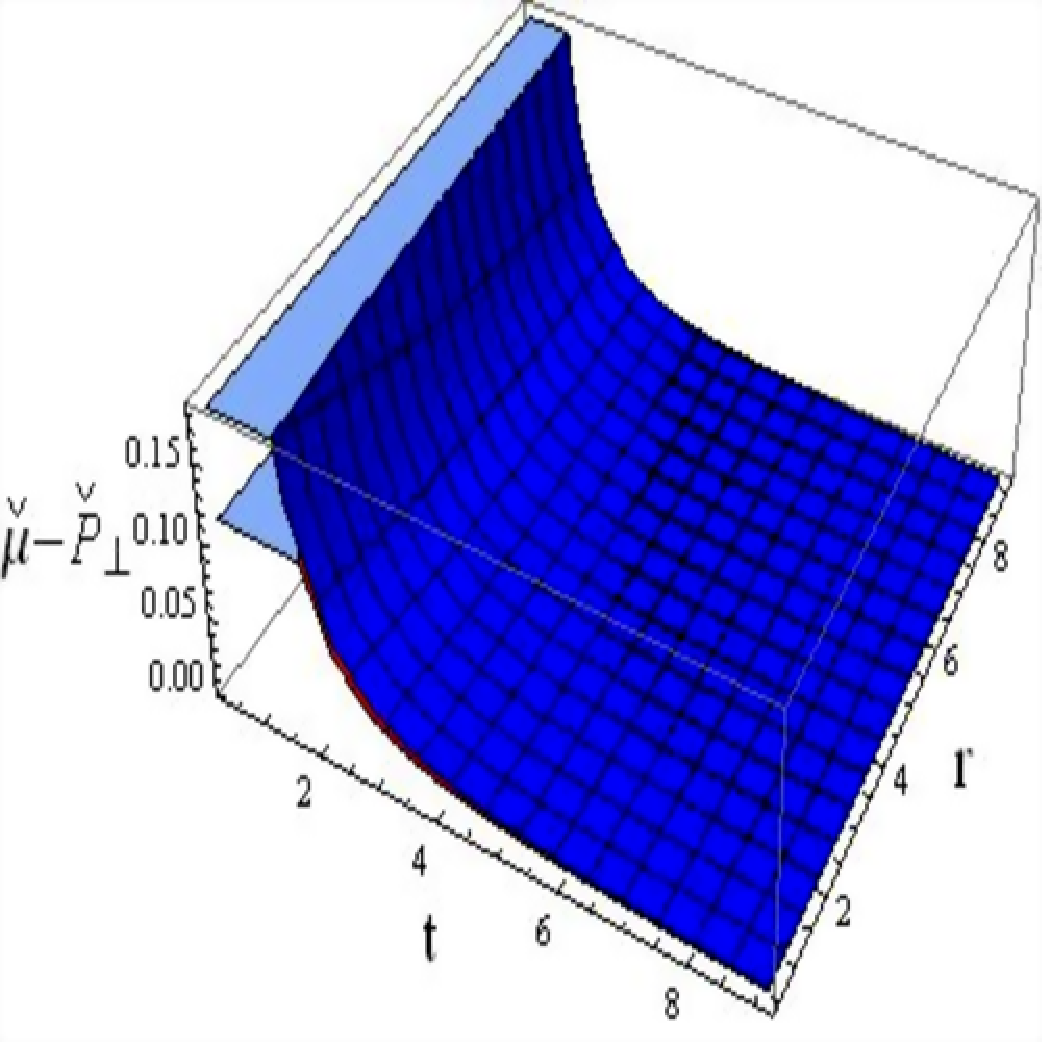,width=0.4\linewidth}
\epsfig{file=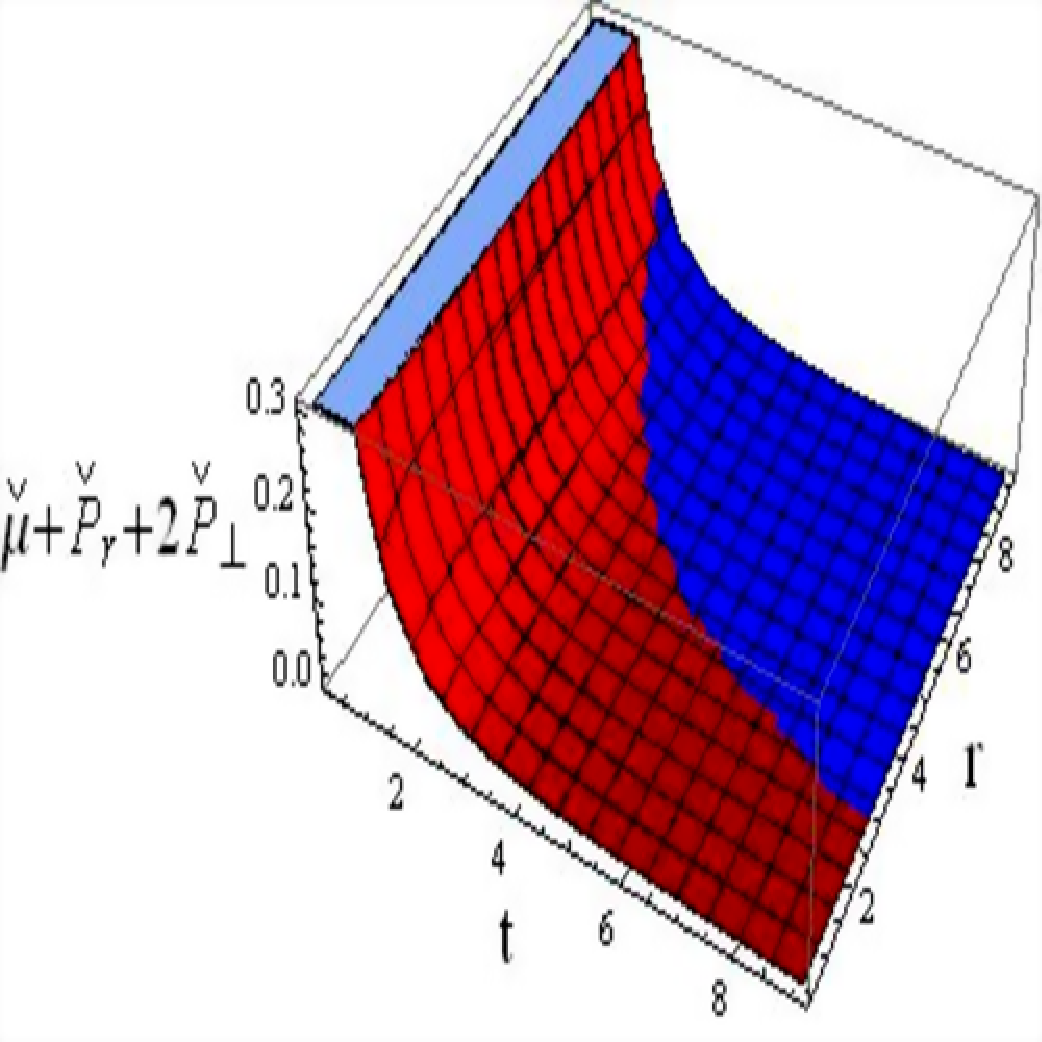,width=0.4\linewidth} \caption{Plots of
$\mathbb{EC}s$ (in km$^{-2}$) with $\gamma_3=0$ for $\varpi=0.1$
(blue) and $0.7$ (red).}
\end{figure}
\begin{figure}\center
\epsfig{file=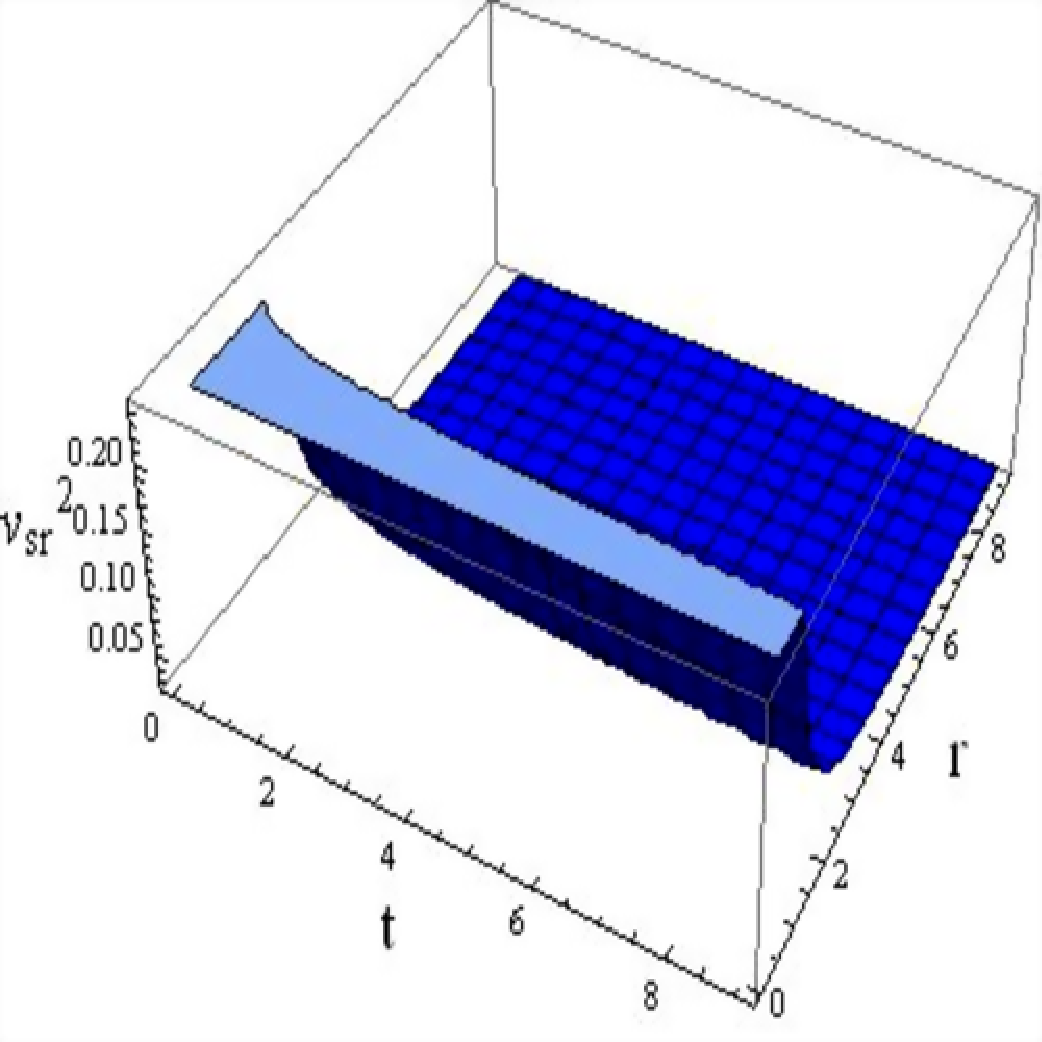,width=0.4\linewidth}\epsfig{file=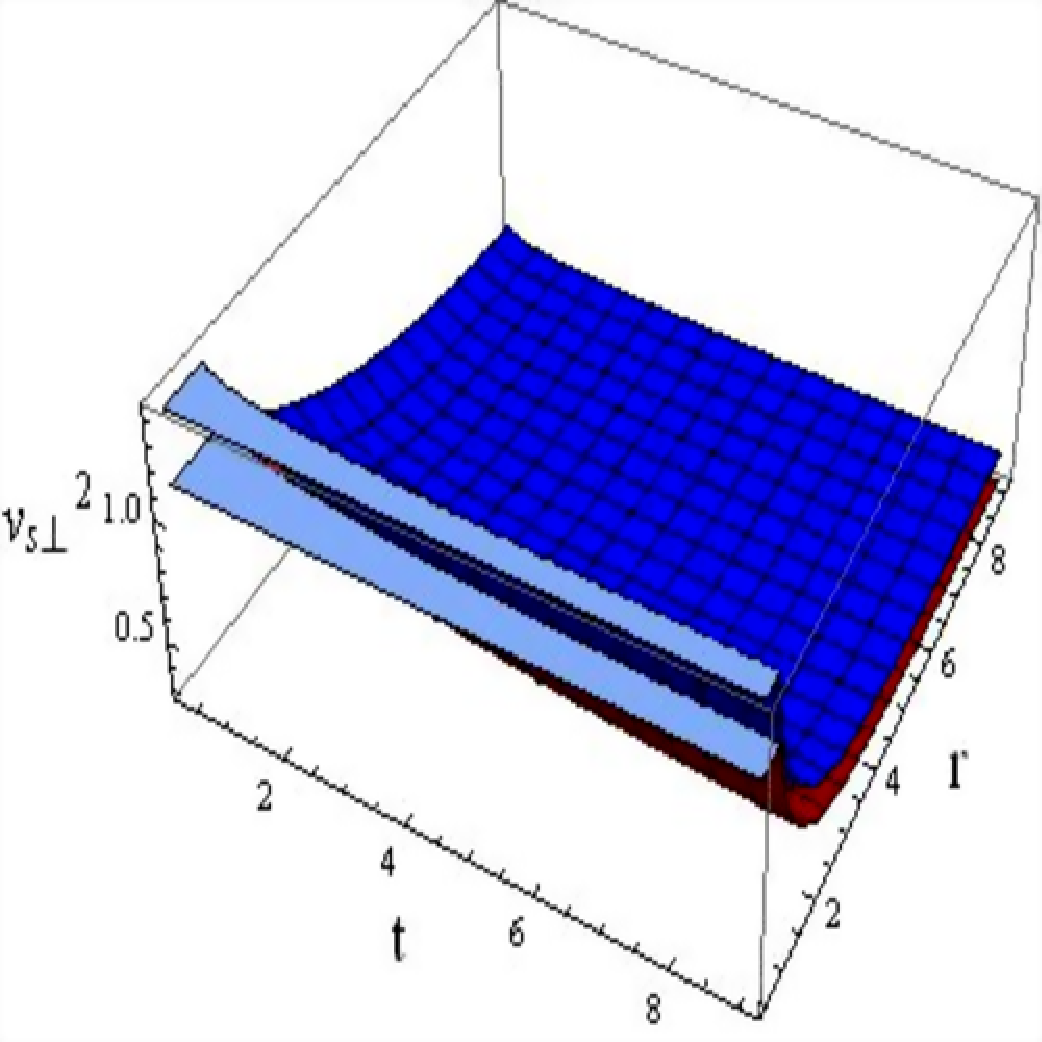,width=0.4\linewidth}
\epsfig{file=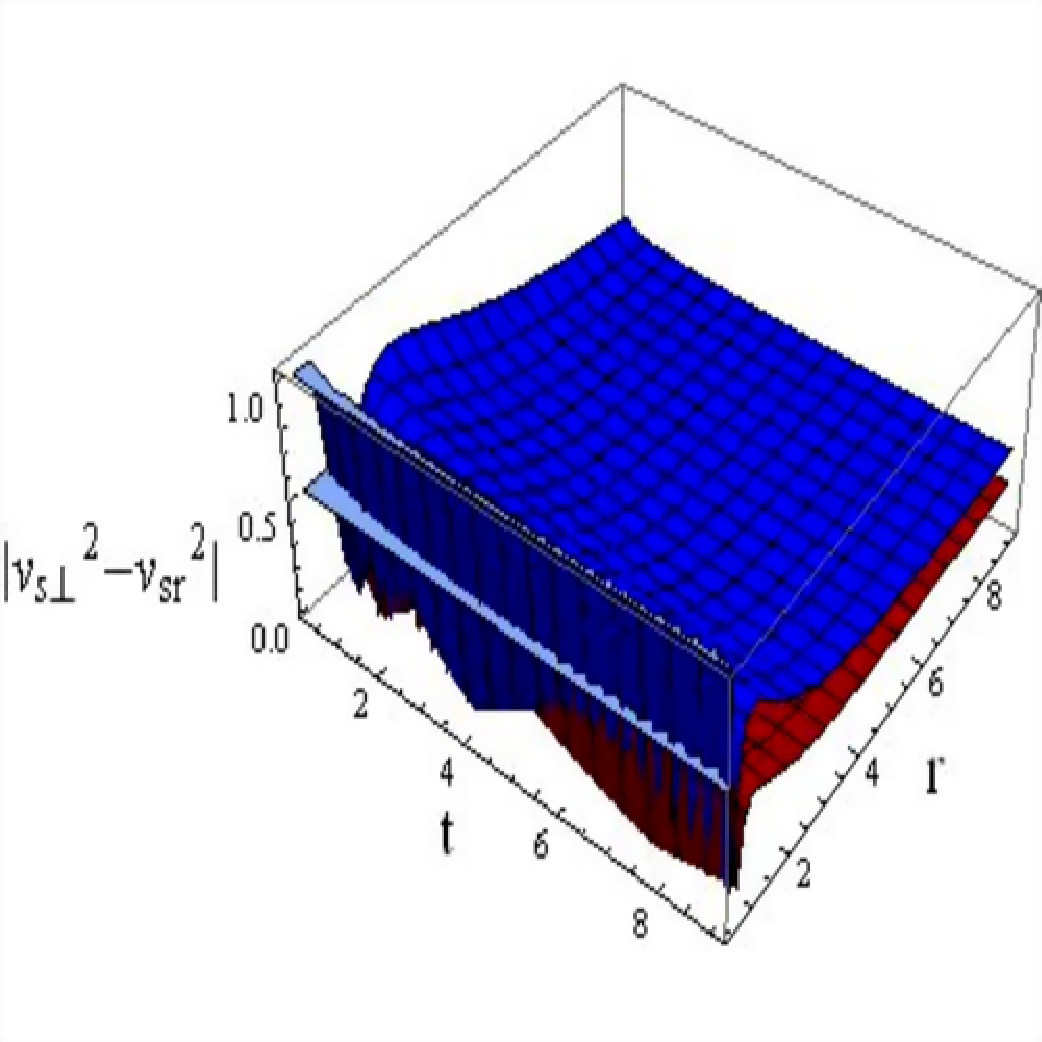,width=0.4\linewidth} \caption{Plots of
$v^{2}_{sr}$, $v^{2}_{s\bot}$ and $|v^{2}_{s\bot}-v^{2}_{sr}|$ with
$\gamma_3=0$ for $\varpi=0.1$ (blue) and $0.7$ (red).}
\end{figure}

\subsection{Vacuum Energy Dominated Phase}

This is the last epoch among different phases of the evolutionary
universe that can be studied by choosing $\gamma_3=-1$ in
$\mathbb{E}o\mathbb{S}$ \eqref{16a}. Since the whole matter density
is dominated by the vacuum (also known as dark energy), thus
accelerating expansion has been observed in this era. The behavior
of state determinants is observed in Figure \textbf{8}. The
decreasing profile of density is obtained with time as well as
increasing $\varpi$, representing expanding universe that is
consistent with previously discussed eras (upper left). We also
observe that the expansion rate increases, as a repulsive force
exists due to the negative profile of radial/tangential pressures.
Furthermore, this rate becomes faster with the increase in the
parameter $\varpi$. Figure \textbf{9} demonstrates that the weak,
strong and null $\mathbb{EC}s$ are not satisfied leading to an
unviable model. The graphs in Figure \textbf{10} are inconsistent
with the respective stability criteria for the considered values of
$\varpi$, hence, the extended solution for $\gamma_3=-1$ is
unstable.
\begin{figure}\center
\epsfig{file=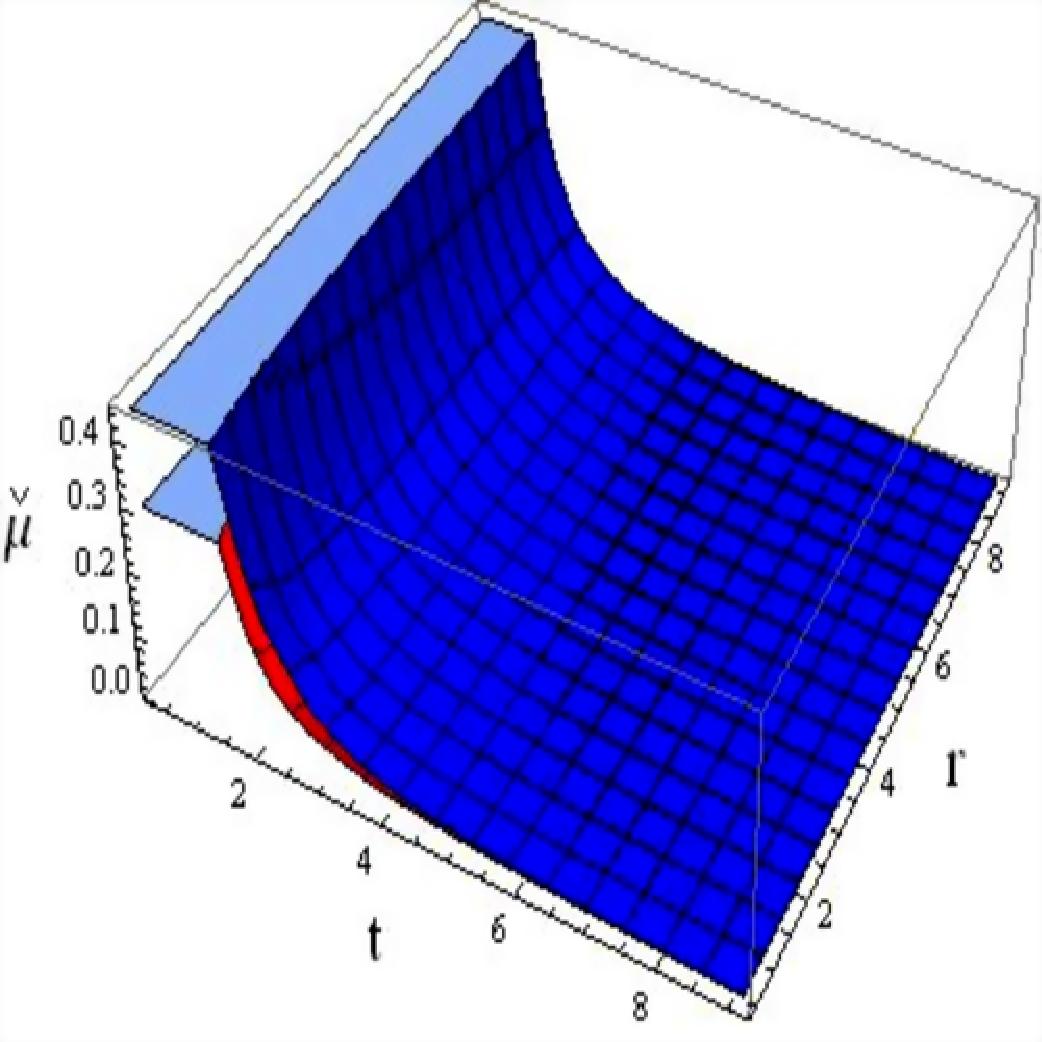,width=0.4\linewidth}\epsfig{file=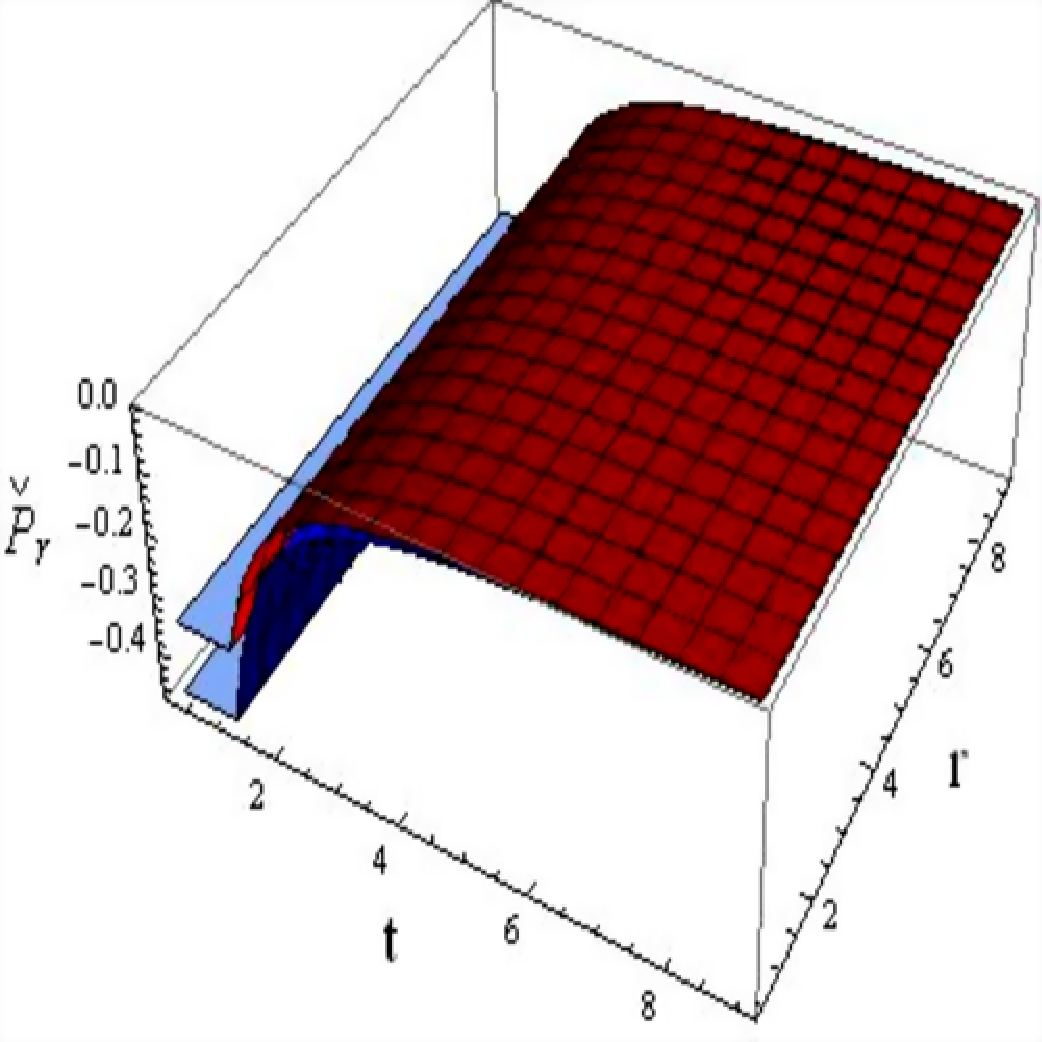,width=0.4\linewidth}
\epsfig{file=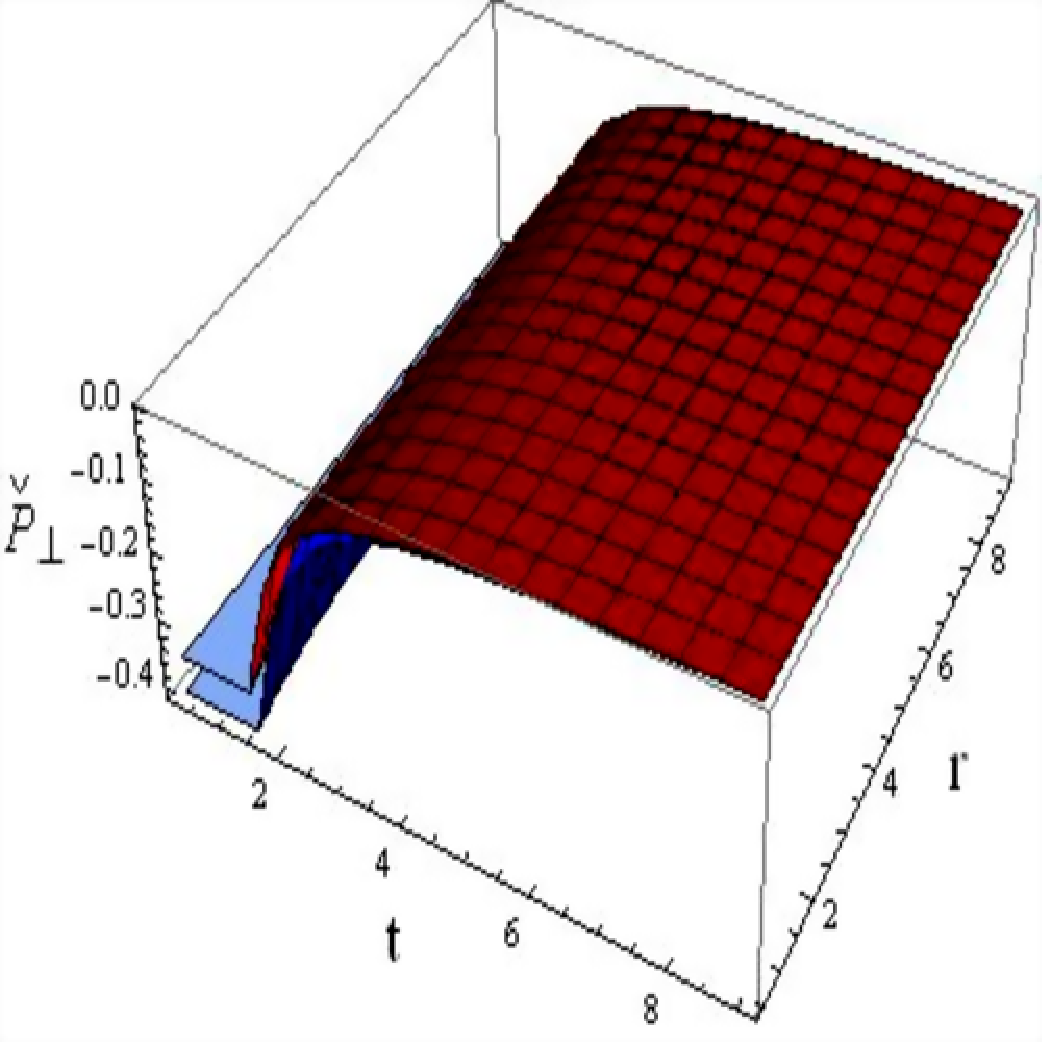,width=0.4\linewidth}\epsfig{file=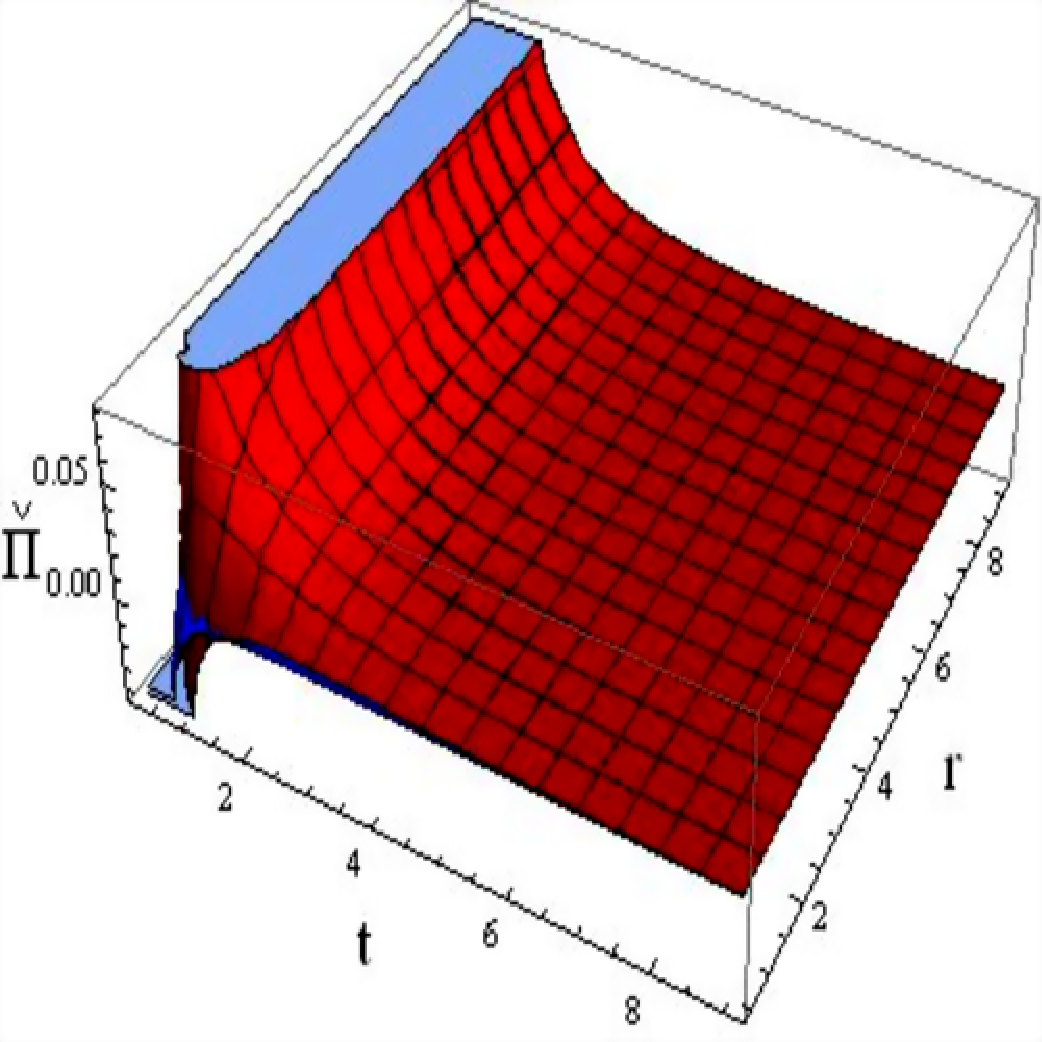,width=0.4\linewidth}
\caption{Plots of matter determinants and anisotropy (in km$^{-2}$)
with $\gamma_3=-1$ for $\varpi=0.1$ (blue) and $0.7$ (red).}
\end{figure}
\begin{figure}\center
\epsfig{file=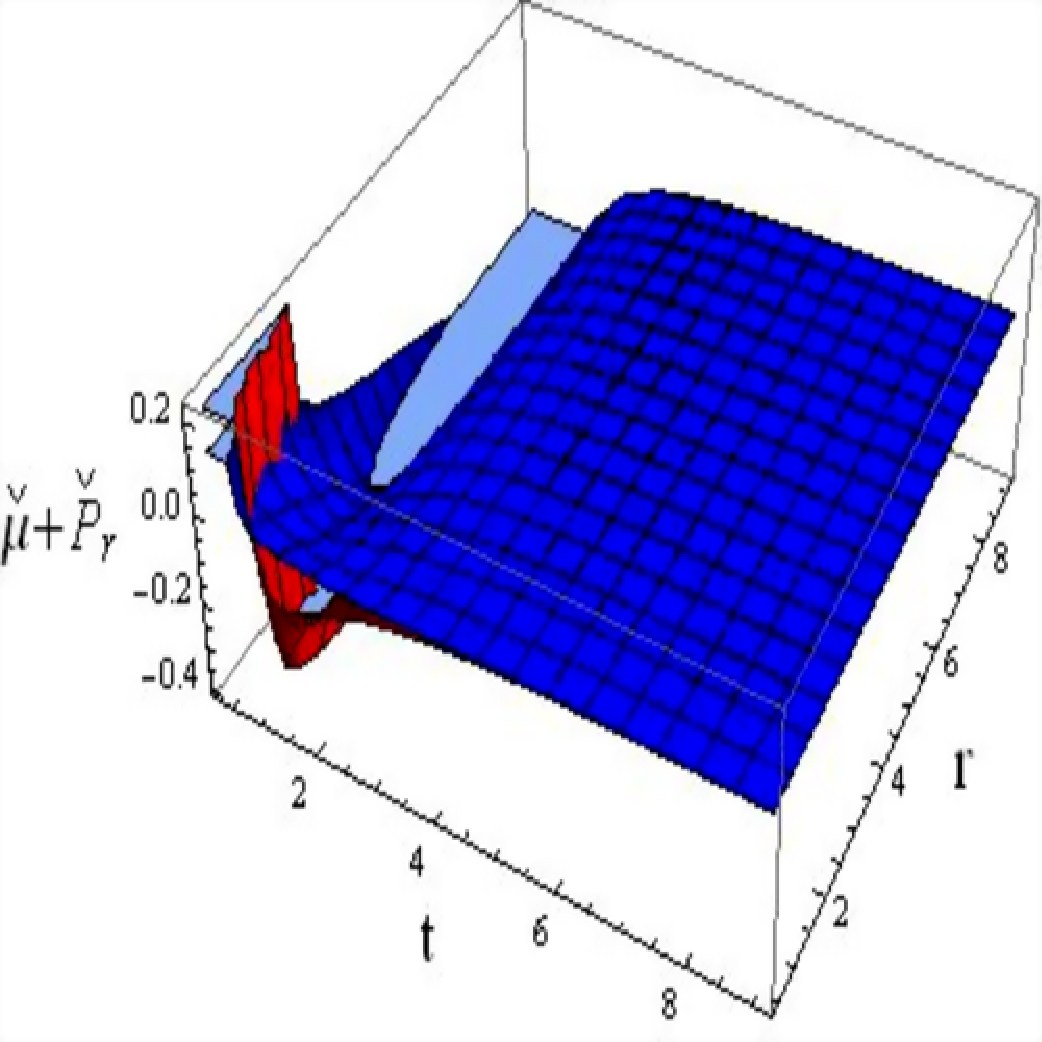,width=0.4\linewidth}\epsfig{file=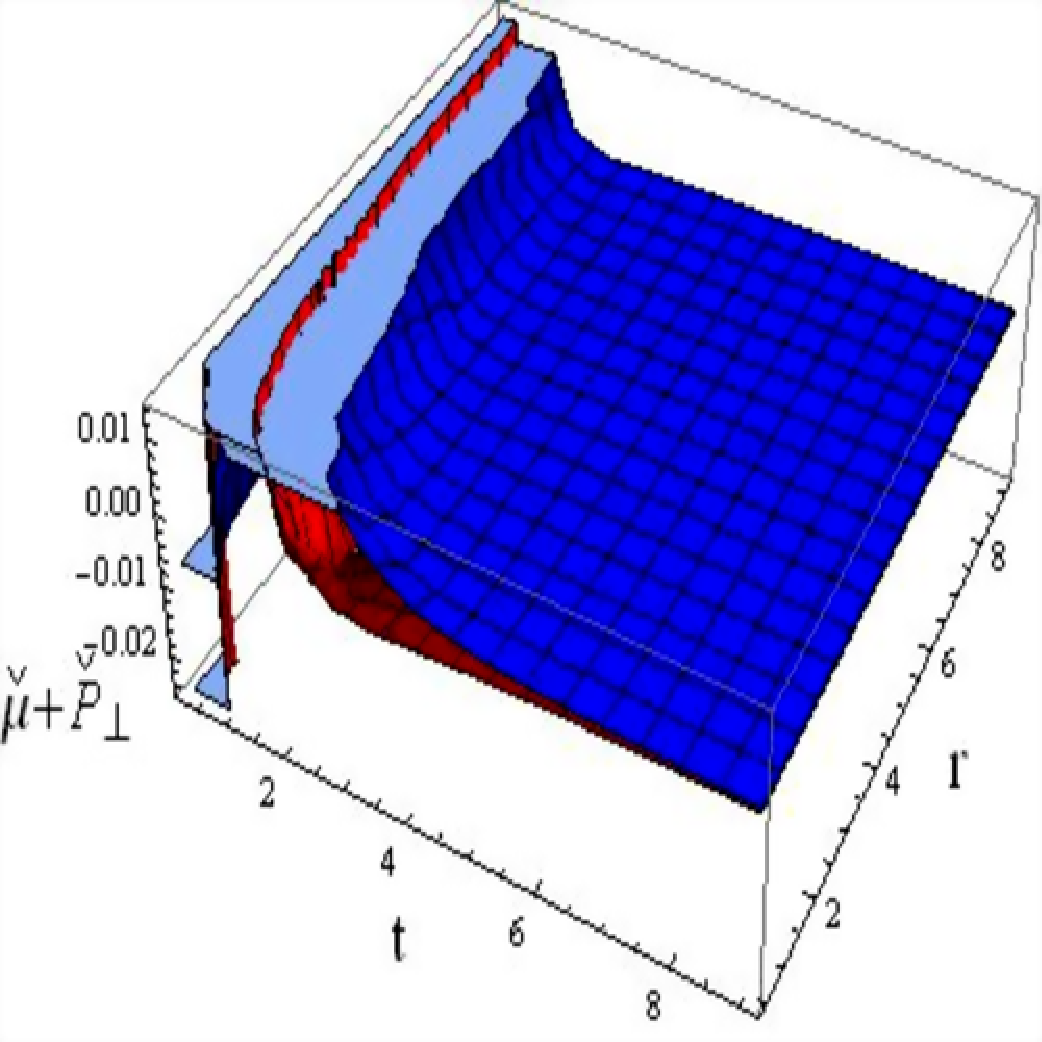,width=0.4\linewidth}
\epsfig{file=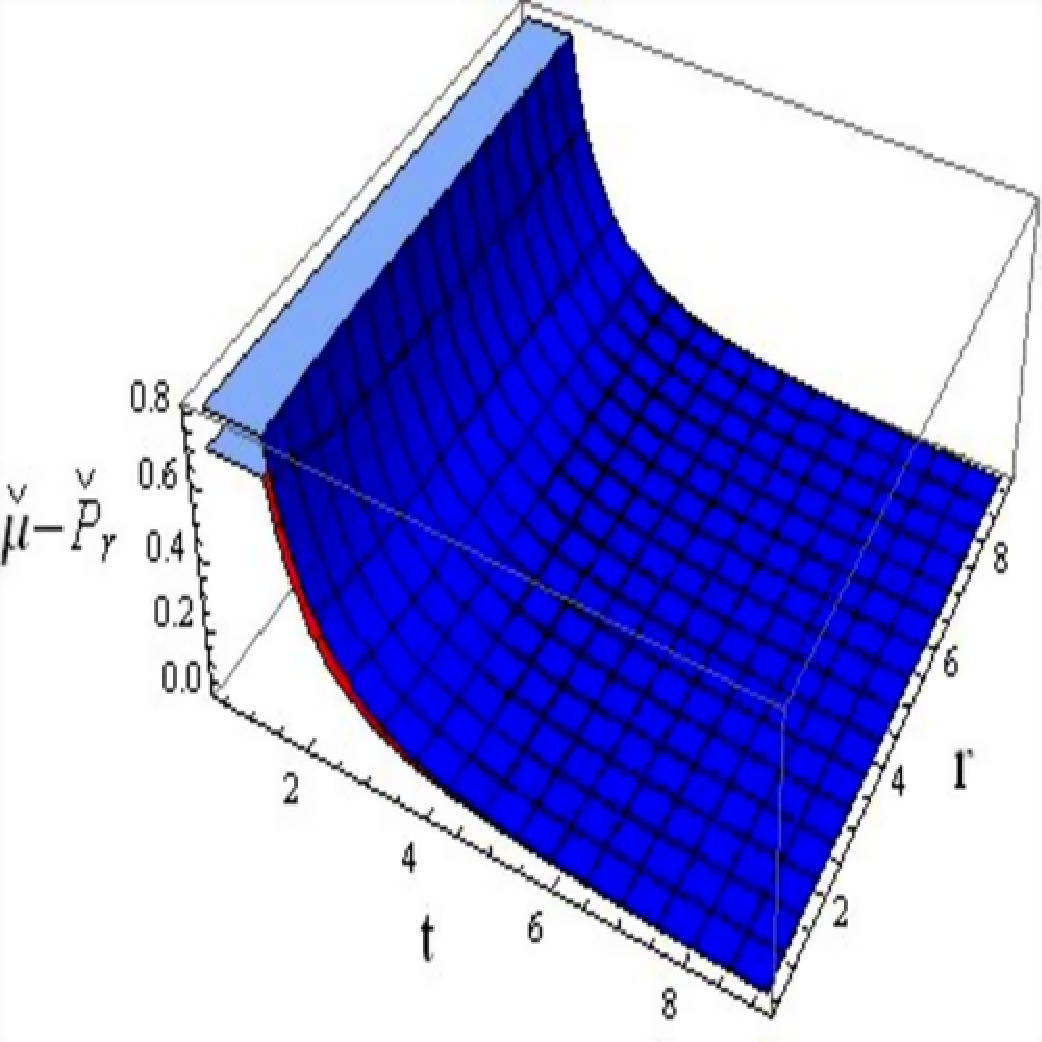,width=0.4\linewidth}\epsfig{file=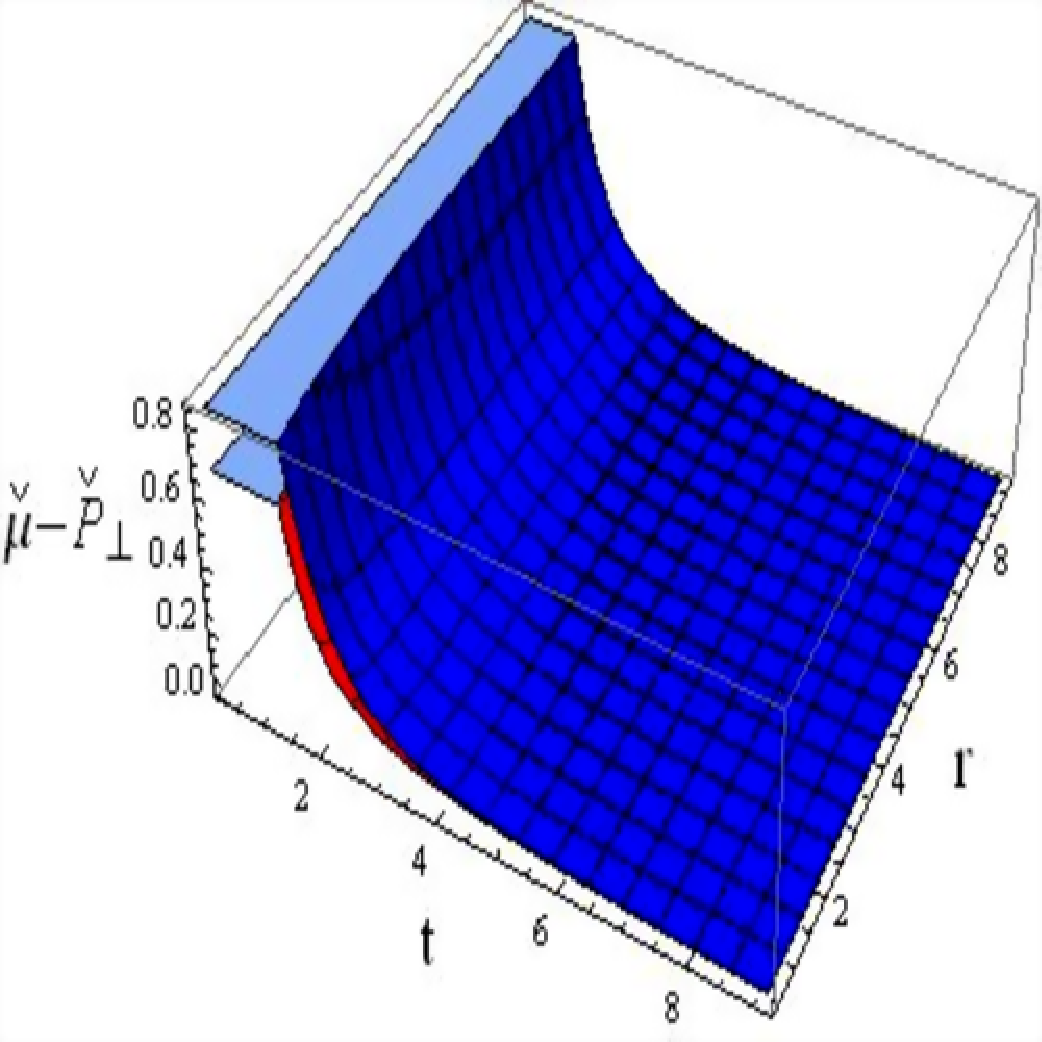,width=0.4\linewidth}
\epsfig{file=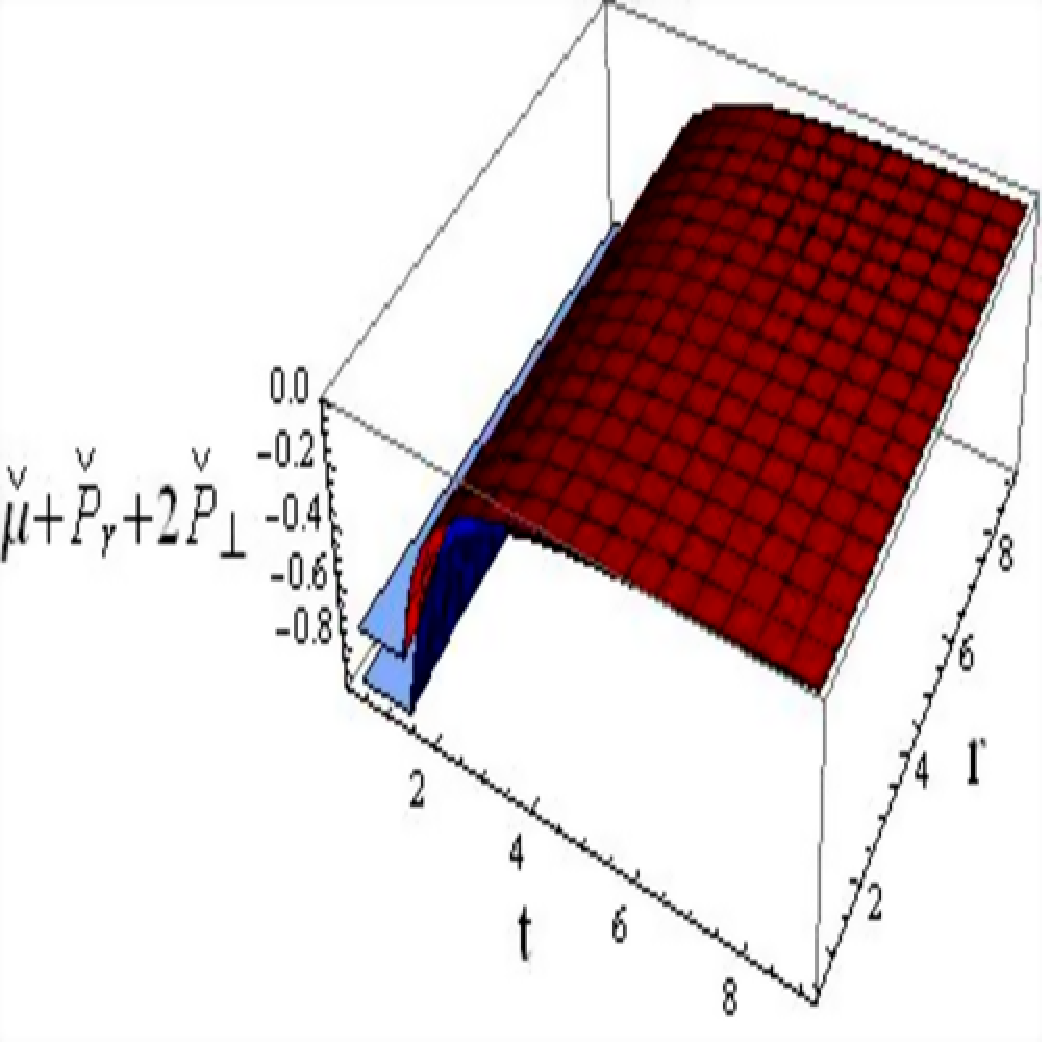,width=0.4\linewidth} \caption{Plots of
$\mathbb{EC}s$ (in km$^{-2}$) with $\gamma_3=-1$ for $\varpi=0.1$
(blue) and $0.7$ (red).}
\end{figure}
\begin{figure}\center
\epsfig{file=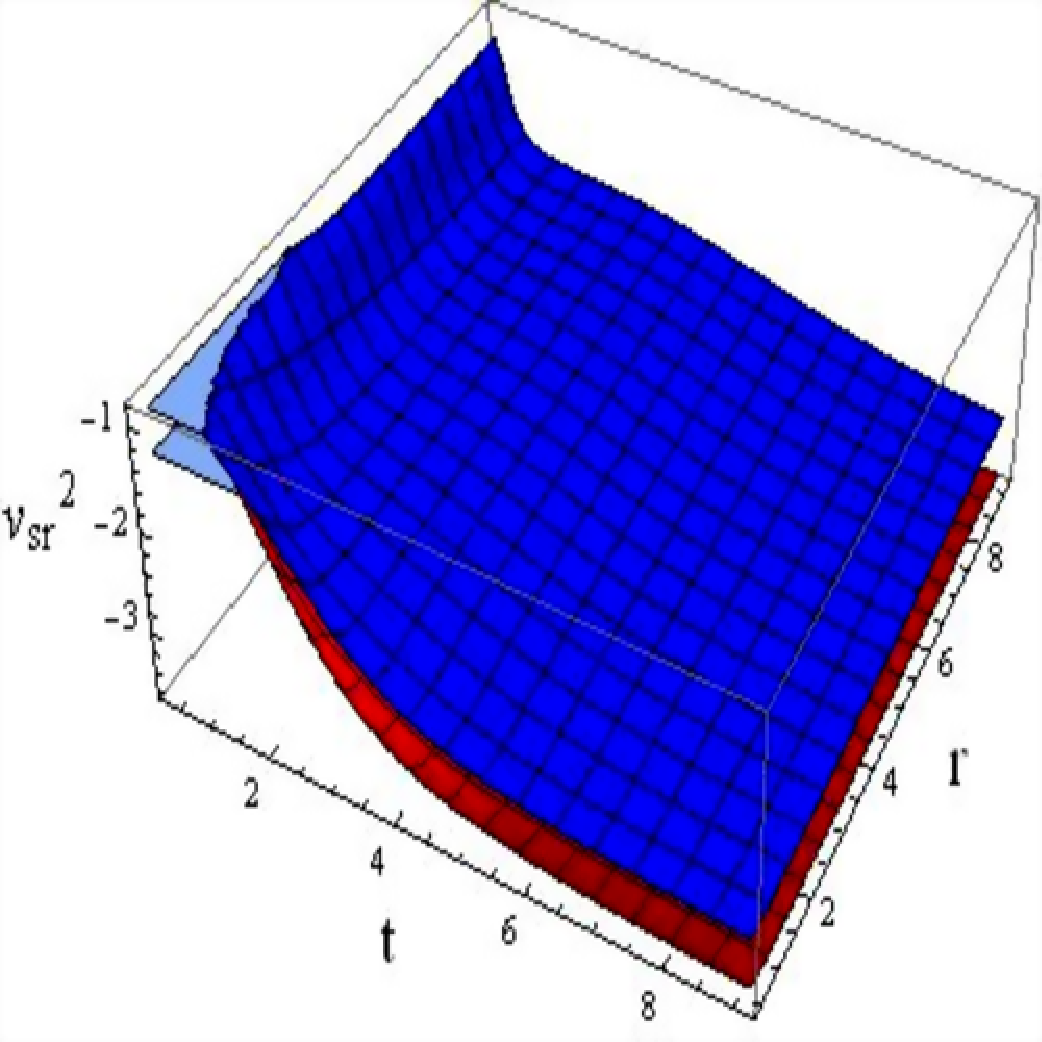,width=0.4\linewidth}\epsfig{file=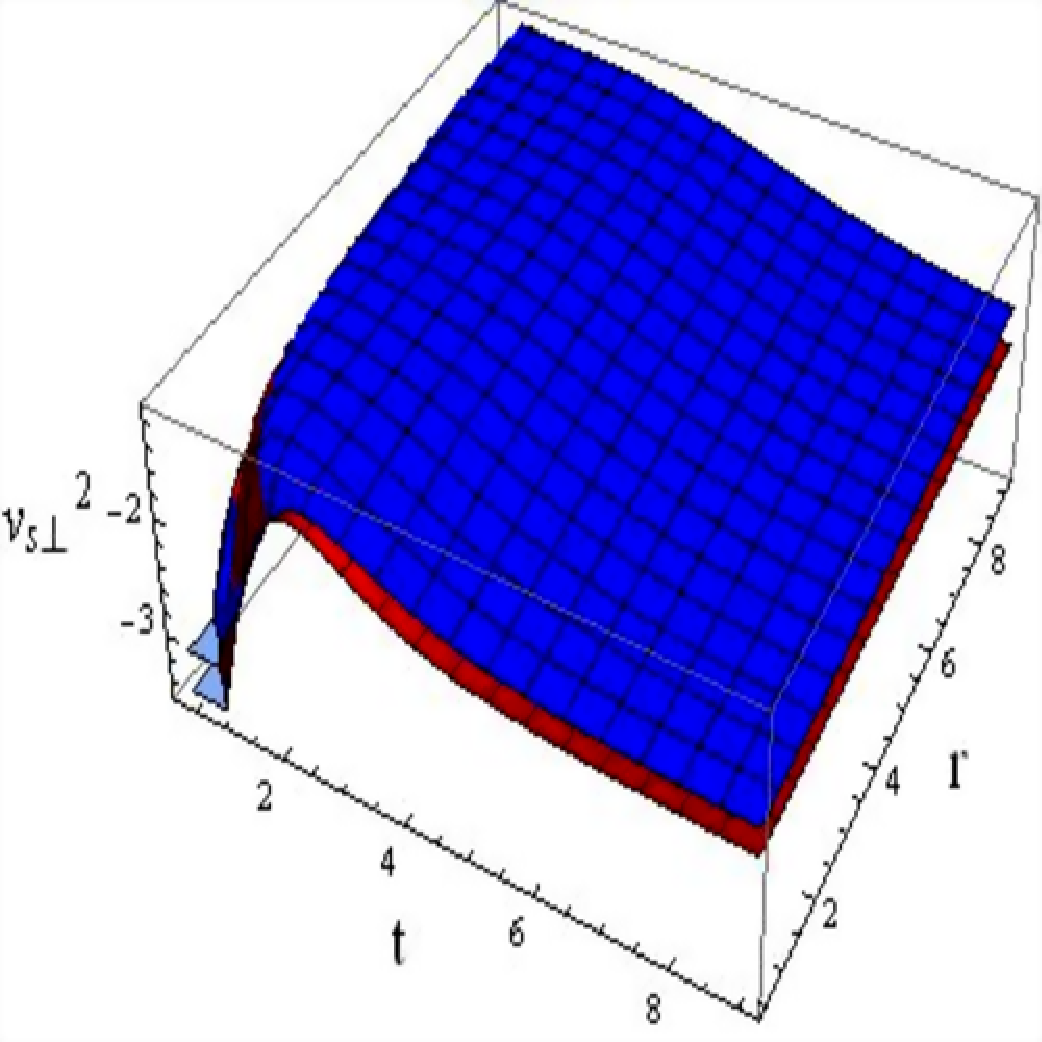,width=0.4\linewidth}
\epsfig{file=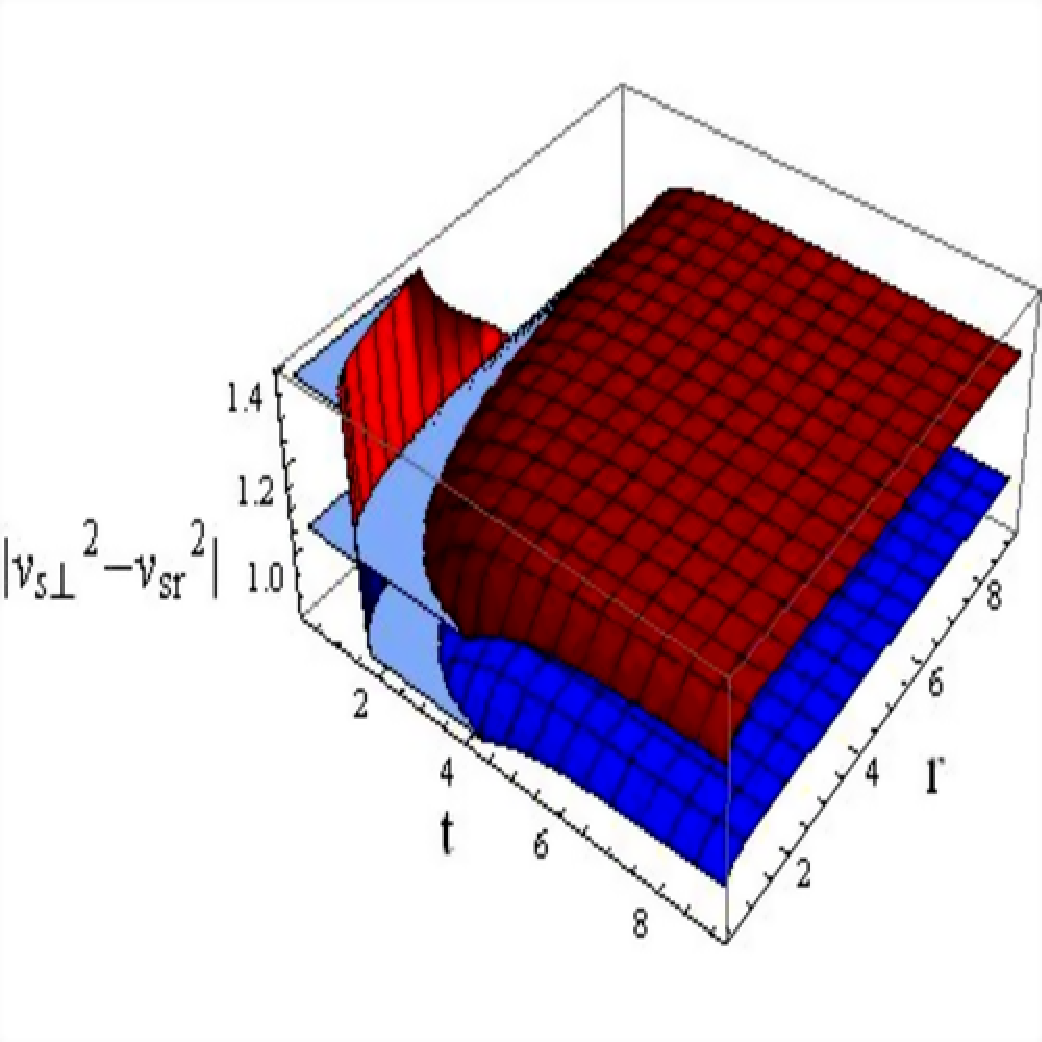,width=0.4\linewidth} \caption{Plots of
$v^{2}_{sr}$, $v^{2}_{s\bot}$ and $|v^{2}_{s\bot}-v^{2}_{sr}|$ with
$\gamma_3=-1$ for $\varpi=0.1$ (blue) and $0.7$ (red).}
\end{figure}

\section{Conclusions}

This paper is based on the formulation of anisotropic extension to
the existing isotropic solution corresponding to a non-static sphere
in the background of
$f(\mathcal{R},\mathcal{T},\mathcal{R}_{\mathrm{a}\mathrm{b}}\mathcal{T}^{\mathrm{a}\mathrm{b}})$
theory. In order to do this, we adopt a decoupling approach through
MGD along with a standard non-minimal model as
$\mathcal{R}+2\gamma_1\mathcal{T}
+\gamma_2\mathcal{R}_{\mathrm{a}\mathrm{b}}\mathcal{T}^{\mathrm{a}\mathrm{b}}$
to get some meaningful results. Since the interior of the seed
source is filled with isotropic fluid, therefore we have added
another source to include the effect of anisotropy in the system
which is being controlled by a parameter $\varpi$. After
implementing the transformation only on $g_{rr}$ function in the
field equations, we have separated them into isotropic and
anisotropic arrays for $\varpi=0$ and $1$, respectively. The
solution to an isotropic sector (characterized by a flat FLRW
metric) has been developed by choosing a particular form of the
scale factor. We have then used a linear $\mathbb{E}o\mathbb{S}$
relating matter variables to discuss different evolutionary phases
of our universe for multiple values of a parameter $\gamma_3$. The
field equations for an additional sector have been solved by
employing a density-like constraint. The combination of both
solutions ultimately provides the required anisotropic model
\eqref{42}-\eqref{42c} whose feasibility has further been explored
for different values of $\gamma_3$ and $\varpi$ along with
$\mathrm{k}=0$. A brief description of the obtained results is given
in the following.
\begin{itemize}
\item The evolutionary phase corresponding to $\gamma_3=\frac{1}{3}$ offers a positive and monotonically decreasing trend
of the pressure components and energy density with time, indicating
the expansion of our universe (Figure \textbf{2}). The formulated
solution, in this case, has shown a viable as well as a stable model
for chosen values of the parameters $\varpi$ and $\gamma_3$ (Figures
\textbf{3,~4}).
\item The profile of state determinants for $\gamma_3=0$ (i.e., matter-dominated phase) has been observed consistent
with the previous model. However, the existence of a more dense
profile is observed in this era as compared to the previous one
(Figure \textbf{5}). This solution is found to be viable but
unstable in the whole domain (Figures \textbf{6,~7}).
\item The energy density \eqref{42} has shown decreasing trend with the increase in time for $\gamma_3=-1$.
On the other hand, we observe the presence of a robust force with
immense repulsive nature due to the negative profile of both
pressure ingredients (Figure \textbf{8}). The viability and
stability criteria have not been fulfilled, hence the vacuum energy
dominated phase does not provide an acceptable model (Figures
\textbf{9,~10}).
\end{itemize}

It is worth noting here that the decoupling parameter $\varpi$ is in
direct relation with pressure components and has an inverse
relationship with the energy density in the above evolutionary
epochs. We conclude that our results in
$f(\mathcal{R},\mathcal{T},\mathcal{R}_{\mathrm{a}\mathrm{b}}\mathcal{T}^{\mathrm{a}\mathrm{b}})$
scenario are consistent with Brans-Dicke gravity, as only the
radiation-dominated era provides stable results in both scenarios
\cite{16a}. However, this theory offers less efficient results in
comparison with $f(\mathcal{R},\mathcal{T})$ gravity. The results of
this paper can be reduced in $\mathbb{GR}$ only for
$\gamma_1=0=\gamma_2$.

\section*{Appendix A}

The entities multiplied by $\gamma_2$ in the field equations
\eqref{8}-\eqref{9b} are
\begin{align}\nonumber
\mathcal{T}_{0}^{0(D)}&=\mu\bigg\{e^{-\alpha_1}\bigg(\frac{3\dot{\alpha}_1^2}{2}+\frac{\dot{\alpha}_1\alpha_2'}{4}
+\frac{3\dot{\mathcal{C}}\dot{\alpha}_1}{\mathcal{C}}-\frac{\dot{\alpha}_2^2}{2}+\frac{\dot{\alpha}_1\dot{\alpha}_2}{2}
-\frac{4\ddot{\mathcal{C}}}{\mathcal{C}}-\ddot{\alpha}_2\bigg)+\frac{\mathcal{R}}{2}\\\nonumber
&+e^{-\alpha_2}\bigg(\frac{\mathcal{C}'\alpha_1'}{\mathcal{C}}-\frac{\alpha_1'\alpha_2'}{4}\bigg)\bigg\}
+\mu'e^{-\alpha_2}\bigg(\frac{\alpha_2'}{4}-\frac{\mathcal{C}'}{\mathcal{C}}-\alpha_1'\bigg)
+\dot{\mu}\dot{\mathcal{C}}e^{-\alpha_1}\\\nonumber
&-\frac{\mu''e^{-\alpha_2}}{2}+P\bigg\{e^{-\alpha_2}\bigg(\frac{3\alpha_1'^2}{2}+\frac{\alpha_1''}{2}
-\frac{2\mathcal{C}'^2}{\mathcal{C}^2}\bigg)+e^{-\alpha_1}\bigg(\frac{\dot{\alpha}_2^2}{4}
+\frac{2\dot{\mathcal{C}}^2}{\mathcal{C}^2}\bigg)\bigg\}\\\nonumber
&+P'e^{-\alpha_2}\bigg(\frac{\mathcal{C}'}{\mathcal{C}}-\frac{5\alpha_2'}{4}\bigg)
-\dot{P}e^{-\alpha_1}\bigg(\frac{\dot{\mathcal{C}}}{\mathcal{C}}+\frac{\dot{\alpha}_2}{4}\bigg)+\frac{P''e^{-\alpha_2}}{2}
-\frac{\mathcal{Q}}{2},\\\nonumber
\mathcal{T}_{1}^{1(D)}&=\mu\bigg\{e^{-\alpha_1}\bigg(\frac{\ddot{\alpha}_1}{2}-\frac{5\dot{\alpha}_1^2}{4}\bigg)
+\frac{\alpha_1\alpha_1'^2e^{-\alpha_1-\alpha_2}}{4}\bigg\}+\frac{\mu'\alpha_1\alpha_1'e^{-\alpha_1-\alpha_2}}{4}\\\nonumber
&+\frac{5\dot{\mu}\dot{\alpha}_1e^{-\alpha_1}}{4}-\frac{\ddot{\mu}e^{-\alpha_1}}{2}
+P\bigg\{e^{-\alpha_1}\bigg(\frac{\dot{\alpha}_1\dot{\alpha}_2}{4}-\frac{2\dot{\mathcal{C}}^2}{\mathcal{C}^2}
-\frac{\dot{\mathcal{C}}\dot{\alpha}_2}{\mathcal{C}}-\frac{\ddot{\alpha}_2}{2}\bigg)\\\nonumber
&+e^{-\alpha_2}\bigg(\frac{2\mathcal{C}'^2}{\mathcal{C}^2}-\frac{3\alpha_2'^2}{2}-\frac{\mathcal{C}'\alpha_2'}{\mathcal{C}}
+\frac{\alpha_1'^2}{2}-\frac{\alpha_1'\alpha_2'}{2}+\frac{4\ddot{\mathcal{C}}}{\mathcal{C}}+\alpha_1''\bigg)
+\frac{\mathcal{R}}{2}\\\nonumber
&+\frac{\alpha_1\alpha_1'\alpha_2'e^{-\alpha_1-\alpha_2}}{4}\bigg\}+P'\bigg\{e^{-\alpha_2}
\bigg(\frac{2\mathcal{C}'}{\mathcal{C}}+\frac{\alpha_2'}{2}\bigg)
+\frac{\alpha_1\alpha_1'e^{-\alpha_1-\alpha_2}}{4}\bigg\}\\\nonumber
&+\dot{P}e^{-\alpha_1}\bigg(\frac{2\dot{\mathcal{C}}}{\mathcal{C}}-\frac{\dot{\alpha}_1}{4}+\dot{\alpha}_2\bigg)
+\frac{\ddot{P}e^{-\alpha_1}}{2}+\frac{\mathcal{Q}}{2},\\\nonumber
\mathcal{T}_{2}^{2(D)}&=\mu\bigg\{e^{-\alpha_1}\bigg(\frac{\ddot{\alpha}_1}{2}-\frac{5\dot{\alpha}_1^2}{4}\bigg)
+\frac{\alpha_1\alpha_1'^2e^{-\alpha_1-\alpha_2}}{4}\bigg\}+\frac{\mu'\alpha_1\alpha_1'e^{-\alpha_1-\alpha_2}}{4}\\\nonumber
&+\frac{5\dot{\mu}\dot{\alpha}_1e^{-\alpha_1}}{4}-\frac{\ddot{\mu}e^{-\alpha_1}}{2}
+P\bigg\{\frac{\mathcal{R}}{2}-\frac{\alpha_1\alpha_1'C'e^{-\alpha_1-\alpha_2}}{2C}
+e^{-\alpha_2}\bigg(\frac{\mathcal{C}'^2}{\mathcal{C}^2}\\\nonumber
&-\frac{3\alpha_2'^2}{4}+\frac{\mathcal{C}''}{\mathcal{C}}-\frac{\mathcal{C}'\alpha_2'}{2\mathcal{C}}
+\frac{\mathcal{C}'\alpha_1'}{\mathcal{C}}\bigg)+e^{-\alpha_1}\bigg(\frac{\dot{\mathcal{C}}\dot{\alpha}_1}{2\mathcal{C}}
-\frac{\dot{\mathcal{C}}^2}{\mathcal{C}^2}-\frac{\dot{\mathcal{C}}\dot{\alpha}_2}{2\mathcal{C}}
-\frac{\ddot{\mathcal{C}}}{\mathcal{C}}\\\nonumber
&+\frac{\alpha_2''}{2}-\frac{\dot{\alpha}_2^2}{4}\bigg)\bigg\}+P'\bigg\{3e^{-\alpha_2}\bigg(\frac{\alpha_2'}{2}
-\frac{\mathcal{C}'}{\mathcal{C}}\bigg)-\frac{\alpha_1\alpha_1'e^{-\alpha_1-\alpha_2}}{4}\bigg\}\\\nonumber
&+\dot{P}e^{-\alpha_1}\bigg(\frac{3\dot{\mathcal{C}}}{\mathcal{C}}-\frac{\dot{\alpha}_1}{4}
+\frac{\dot{\alpha}_2}{2}\bigg)-P''e^{-\alpha_2}+\frac{\ddot{P}e^{-\alpha_1}}{2}+\frac{\mathcal{Q}}{2},\\\nonumber
\mathcal{T}_{1}^{0(D)}&=e^{-\alpha_1}\bigg\{\mu\bigg(\frac{\dot{\mathcal{C}}\alpha_1'}{\mathcal{C}}
+\frac{\mathcal{C}'\dot{\alpha}_2}{\mathcal{C}}-\frac{2\dot{\mathcal{C}}'}{\mathcal{C}}\bigg)
-P\bigg(\frac{\dot{\mathcal{C}}\alpha_1'}{\mathcal{C}}+\frac{\mathcal{C}'\dot{\alpha}_2}{\mathcal{C}}
-\frac{2\dot{\mathcal{C}}'}{\mathcal{C}}\bigg)\\\nonumber
&+\frac{\dot{\mu}\alpha_1'}{4}+\frac{\mu'\dot{\alpha}_2}{4}-\frac{\dot{\mu}'}{2}-\frac{\dot{P}\alpha_1'}{4}
-\frac{P'\dot{\alpha}_2}{4}+\frac{\dot{P}'}{2}\bigg\}.
\end{align}\\
\textbf{Data Availability Statement:} No new data were created or
analysed in this study.


\begin{thebibliography}{40}

\bibitem{1} Riess, A.G. et al.: Astron. J. \textbf{116}(1998)1009; Perlmutter, S. et al.: Astrophys. J. \textbf{517}(1999)565.

\bibitem{1a} Bennet, C.L. et al.: Astrophys. J. Suppl. \textbf{148}(2003)97.

\bibitem{1b} Allen, S.W., Schmidt, R.W. and Bridle, S.L.: Mon. Not. R. Astr. Soc.
\textbf{346}(2003)593.

\bibitem{2} Capozziello, S. et al.: Class. Quantum Grav. \textbf{25}(2008)
085004; Nojiri, S. et al.: Phys. Lett. B \textbf{681}(2009)74.

\bibitem{2a} de Felice, A. and Tsujikawa, S.: Living Rev.
Relativ. \textbf{13}(2010)3; Nojiri, S. and Odintsov, S.D.: Phys.
Rep. \textbf{505}(2011)59.

\bibitem{9} Sharif, M. and Kausar, H.R.: J. Cosmol. Astropart. Phys.
\textbf{07}(2011)022.

\bibitem{9g} Astashenok, A.V., Capozziello, S. and Odintsov, S.D.: J. Cosmol. Astropart. Phys. \textbf{01}(2015)001;
Phys. Lett. B \textbf{742}(2015)160.

\bibitem{9h} Shamir, M.F. and Fayyaz, I.: Theor. Math. Phys.
\textbf{202}(2020)112; Zubair, M., Saleem, R. and Lodhi, M.: Int. J.
Geom. Methods Mod. \textbf{17}(2020)2050185; Shamir, M.F. and Malik,
A.: Chin. J. Phys. \textbf{69}(2021)312; Pretel, J.M.Z.,
Arba{\~n}il, J.D.V., Duarte, S.B., Jor{\'a}s, S.E. and Reis, R.R.R.:
J. Cosmol. Astropart. Phys. \textbf{09}(2022)058.

\bibitem{10} Bertolami, O., Boehmer, C.G., Harko, T. and Lobo, F.S.N.: Phys. Rev. D \textbf{75}(2007)104016.

\bibitem{20} Harko, T., Lobo, F.S.N., Nojiri, S. and Odintsov, S.D.: Phys. Rev. D \textbf{84}(2011)024020.

\bibitem{22a} Baffou, E.H. et al.: Astrophys. Space Sci.
\textbf{356}(2015)173.

\bibitem{22ab} Sharif, M. and Naseer, T.: Eur. Phys. J. Plus
\textbf{137}(2022)1304; Class. Quantum Gravit.
\textbf{40}(2023)035009.

\bibitem{22} Haghani, Z., Harko, T., Lobo, F.S.N., Sepangi, H.R. and Shahidi, S.: Phys. Rev. D \textbf{88}(2013)044023.

\bibitem{23} Odintsov, S.D. and S{\'a}ez-G{\'o}mez, D.: Phys. Lett. B \textbf{725}(2013)437.

\bibitem{22b} Sharif, M. and Zubair, M.: J. Cosmol. Astropart. Phys. \textbf{11}(2013)042.

\bibitem{22c} Sharif, M. and Zubair, M.: J. High Energy Phys. \textbf{12}(2013)79.

\bibitem{21} Sharif, M. and Naseer, T.: Phys. Scr. \textbf{97}(2022)055004; Pramana \textbf{96}(2022)119;
Fortschritte der Phys. \textbf{71}(2022)2200147; Phys. Scr.
\textbf{97}(2022)125016.

\bibitem{21b} Yousaf, Z., Bhatti, M.Z. and Naseer, T.: Int. J. Mod. Phys. D \textbf{29}(2020)2050061; Ann. Phys.
\textbf{420}(2020)168267; Eur. Phys. J. Plus \textbf{135}(2020)353;
Phys. Dark Universe \textbf{28}(2020)100535; Sharif, M. and Naseer,
T.: Chin. J. Phys. \textbf{77}(2022)2655; Eur. Phys. J. Plus
\textbf{137}(2022)947.

\bibitem{9a} Wu, K.K.S., Lahav, O. and Rees, M.J.: Nature \textbf{397}(1999)225.

\bibitem{10aaaa} Rodrigues, M.E., Houndjo, M.J.S., Saez-Gomez, D. and Rahaman,
F.: Phys. Rev. D \textbf{86}(2012)104059; Rodrigues, M.E., Salako,
I.G., Houndjo, M.J.S. and Tossa, J.: Int. J. Mod. Phys. D
\textbf{23}(2014)1450004; Rodrigues, M.E., Kpadonou, A.V., Rahaman,
F., Oliveira, P.J. and Houndjo, M.J.S.: Astrophys. Space Sci.
\textbf{357}(2015)129.

\bibitem{10aaa} Schwarz, D.J. and Weinhorst, B.: Astron. Astrophys. \textbf{474}(2007)717; Antoniou, I. and Perivolaropoulos, L.:
J. Cosmol. Astropart. Phys. \textbf{12}(2010)012.

\bibitem{10aaaaa} Colin, J., Mohayaee, R., Sarkar, S. and Shafieloo, A.: Mon. Not. R.
Soc. \textbf{414}(2011)264; Javanmardi, B., Porciani, C., Kroupa, P.
and Pflamm-Altenburg, J.: Astrophys. J. \textbf{810}(2015)47.

\bibitem{10aa} Tiwari, P. and Nusser, A.: J. Cosmol. Astropart. Phys.
\textbf{3}(2016)062.

\bibitem{10ab} Yoon, M. et al.: Mon. Not. R. Soc. \textbf{445}(2014)L60.

\bibitem{10ac} {\v{R}}{\'\i}pa, J. and Shafieloo, A.: Astrophys. J.
\textbf{851}(2017)15.

\bibitem{10a} Migkas, K. and Reiprich, T.H.: Astron. Astrophys. \textbf{611}(2018)A50.

\bibitem{10b} Migkas, K. et al.: Astron. Astrophys. \textbf{636}(2020)A15.

\bibitem{10bc} Sharif, M. and Naseer, T.: Ann. Phys.
\textbf{453}(2023)169311.

\bibitem{11} Ovalle, J.: Phys. Rev. D \textbf{95}(2017)104019.

\bibitem{13} Ovalle, J., Casadio, R., da Rocha, R. and Sotomayor, A.: Eur. Phys. J. C \textbf{78}(2018)122.

\bibitem{14} Gabbanelli, L., Rinc{\'o}n, {\'A}. and Rubio, C.: Eur. Phys. J. C \textbf{78}(2018)370.

\bibitem{17} Estrada, M. and Tello-Ortiz, F.: Eur. Phys. J. Plus \textbf{133}(2018)453.

\bibitem{15} Sharif, M. and Sadiq, S.: Eur. Phys. J. C \textbf{78}(2018)410.

\bibitem{19} Hensh, S. and Stuchl{\'\i}k, Z.: Eur. Phys. J. C \textbf{79}(2019)834.

\bibitem{16} Sharif, M. and Waseem, A.: Ann. Phys. \textbf{405}(2019)14;
Sharif, M. and Saba, S.: Chin. J. Phys. \textbf{59}(2019)481.

\bibitem{19a} Cede{\~n}o, F.X.L and Contreras, E.: Phys. Dark Universe \textbf{28}(2020)100543.

\bibitem{16b} Maurya, S.K. et al.: Phys. Dark Universe \textbf{30}(2020)100640.

\bibitem{16c} Azmat, H. and Zubair, M.: Eur. Phys. J. Plus \textbf{136}(2021)112.

\bibitem{21a} Sharif, M. and Naseer, T.: Chin. J. Phys.
\textbf{73}(2021)179; Int. J. Mod. Phys. D \textbf{31}(2022)2240017;
Indian J. Phys. \textbf{96}(2022)4373; Naseer, T and Sharif, M.:
Universe \textbf{8}(2022)62.

\bibitem{16a} Sharif, M. and Majid, A.: Phys. Scr. \textbf{96}(2021)045003.

\bibitem{39} Shamir, M.F.: Eur. Phys. J. C \textbf{75}(2015)354; Moraes, P.H.R.S., Correa R.A.C. and Lobato, R.V.:
J. Cosmol. Astropart. Phys. \textbf{07}(2017)029.

\bibitem{24ab} Perlmutter, S. et al.: Nature \textbf{391}(1998)51.

\bibitem{24a} Abreu, H., Hernandez, H. and Nunez, L.A.: Class. Quantum Gravit. \textbf{24}(2007)4631.

\bibitem{24} Herrera, L.: Phys. Lett. A \textbf{165}(1992)206.
\end{thebibliography}
\end{document}